\newif\ifmainmode         \mainmodetrue
\newif\ifsuppmode         \suppmodetrue
\newif\ifcombinedmode \combinedmodetrue

\documentclass[
aps,
prx,
showpacs,
preprintnumbers,
twocolumn,
superscriptaddress,
10pt,
longbibliography
]{revtex4-2}

\usepackage[T1]{fontenc}
\usepackage{newtxtext}
\usepackage{newtxmath}

\usepackage[normalem]{ulem}

\usepackage{enumitem}

\usepackage{amsmath}
\usepackage{algorithm}
\usepackage{algpseudocode}
\usepackage{float}
\usepackage{braket}

\usepackage{physics}
\usepackage{amsmath}
\usepackage{mathtools}
\usepackage{dsfont}

\usepackage{appendix}

\usepackage{placeins}

\usepackage{color}
\usepackage[dvipsnames,table]{xcolor}

\definecolor{nqdcolor}{rgb}{0.5586, 0.0586, 0.4219}
\newcommand*{\nqdcolor}{\color{nqdcolor}}

\usepackage{hyperref}
\hypersetup{colorlinks,citecolor=nqdcolor,linkcolor=nqdcolor,urlcolor=nqdcolor}

\usepackage{makecell}
\usepackage{colortbl} 

\usepackage{graphicx}
\usepackage[all]{hypcap}
\graphicspath{{./visual_elements/}}

\usepackage[T1]{fontenc}
\usepackage[utf8]{inputenc}
\usepackage{booktabs}
\usepackage{tabularx}
\usepackage{array}

\newcommand{\affA}{State Key Laboratory of Artificial Microstructure and Mesoscopic Physics, School of Physics, Peking University, Beijing 100871, China}
\newcommand{\affB}{Max Planck Institute for the Physics of Complex Systems, N\"othnitzer Str.~38, 01187 Dresden, Germany.}
\newcommand{\affC}{Theoretical Physics III, Center for Electronic Correlations and Magnetism,
		Institute of Physics, University of Augsburg, 86135 Augsburg, Germany.}
\newcommand{\affD}{Department of Physics, National University of Singapore, Singapore 117542, Singapore}

\newcommand{\nocontentsline}[3]{}
\let\origcontentsline\addcontentsline
\newcommand\stoptoc{\let\addcontentsline\nocontentsline}
\newcommand\resumetoc{\let\addcontentsline\origcontentsline}

\makeatletter
\@ifundefined{ifmainmode}{\newif\ifmainmode\mainmodetrue}{}%
\@ifundefined{ifsuppmode}{\newif\ifsuppmode\suppmodetrue}{}%
\@ifundefined{ifcombinedmode}{\newif\ifcombinedmode\combinedmodetrue}{}%
\makeatother

\usepackage{etoolbox}

\begin{document}

\title{Reinforcement Learning to Harness Approximation Errors for Long-Time Quantum Simulation}
\author{Yu-Bo Shi}
 \affiliation{\affA}
  \affiliation{\affD}
	\author{Markus Heyl}
 \affiliation{\affC}
	\author{Roderich Moessner}
 \affiliation{\affB}
  \author{Marin Bukov}
   \email{mgbukov@pks.mpg.de}
 \affiliation{\affB}
\author{Hongzheng Zhao}
\email{hzhao@pku.edu.cn}
 \affiliation{\affA}
 	\date{\today}
\ifmainmode

\begin{abstract}
Accurate digital quantum simulation at long times is limited by the accumulation of errors inherent to approximate simulation. Here we introduce RL-Trotter, a reinforcement-learning framework that treats unavoidable approximation errors as resources for error correction rather than merely imperfections to suppress. We show that low-dimensional information from conservation laws, such as the energy and energy variance, provides a sufficient learning signal to guide the agent, which learns to adapt a single scalar—the next Trotter step size—without access to the target wave function. By optimizing the entire long-time evolution rather than individual steps, RL-Trotter discovers self-correcting sequences in which later errors compensate for those accumulated earlier, increasing the  accuracy of the long-time dynamics. The learned policies are intrinsically robust to measurement noise, substantially reducing measurement overhead. They also generalize to previously unseen, physically similar initial states and transfer from small, classically simulable systems to systems an order of magnitude larger. This enables a practical protocol based on classical pretraining followed by direct deployment or limited fine-tuning on quantum hardware. Our results establish a broader perspective for quantum algorithms: errors in approximate evolution can be orchestrated into resources for accurate and resource-efficient quantum dynamics.
\end{abstract}

\ifcombinedmode

\makeatletter
\let\savedmaketitle\maketitle
\let\savedauthor\author
\let\savedaffiliation\affiliation
\let\savedemail\email
\let\savedthanks\thanks
\makeatother
\else

\fi 

\maketitle

\stoptoc

\begin{figure*}[t!]
	\centering
	\includegraphics[width=0.95\textwidth]{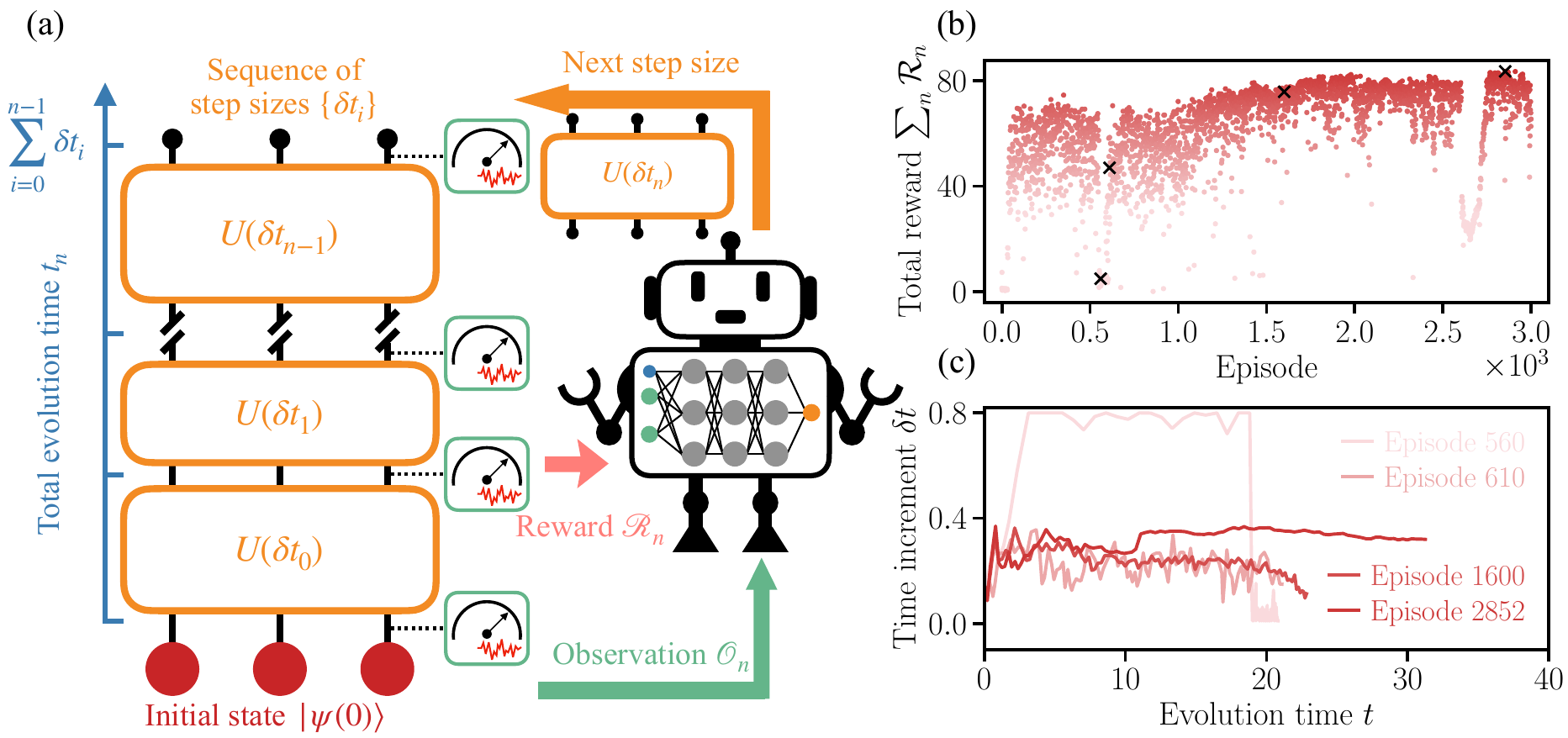}
	\caption{
\textbf{Symmetry-guided RL for adaptive Trotterization.} (a) Closed-loop workflow of RL-Trotter. Starting from the initial state $\lvert\psi(0)\rangle$, the quantum state is propagated through a sequence of Trotter operations $U(\delta t_i)$ defined in Eq.~\eqref{eq.Trotter}, with the accumulated evolution time $t_n=\sum_{i=0}^{n-1}\delta t_i$. After each step, the current noisy measurements of conserved quantities—here, the energy and its variance—together with their initially measured values and $t_n$ are supplied to the RL agent, which selects the next Trotter step size $\delta t_{n}$. 
(b) A representative illustration of the learning process (total episode reward $\sum_n R_n$ versus training episode).
The overall increase and subsequent saturation of the reward indicate progressive learning of an effective Trotter policy. 
Black crosses mark the representative episodes examined in panel (c). (c) Trotter step sizes ${\delta t}$ as a function of evolution time $t$ selected at episodes $560$, $610$, $1600$, and $2852$ (best episode). 
The evolution of these sequences illustrates how training produces a stable adaptive policy that balances the advancement of physical time against violations of the conserved quantities. {See the videos for the training process and Fig.~\ref{quench_dynamics} for the performance of the trained agent.}
}
\label{fig1_illustration}
\end{figure*}

\section{Introduction}

Modern quantum technologies~\cite{rigol2008thermalization,polkovnikov2011colloquium,cheneau2012light} enable the realization of {quantum many-body systems} in experimental platforms as diverse as Rydberg-atom arrays~\cite{bernien2017probing,browaeys2020many}, trapped ions~\cite{zhang2017observation,monroe2021programmable}, and superconducting qubits~\cite{kjaergaard2020superconducting}.
These advances unlock exclusive access to quantum dynamics at system sizes and timescales that are inaccessible to classical computation.
Nevertheless, {simulation} errors~\cite{hauke2012can,poggi2020quantifying} are ubiquitous and unavoidable in quantum devices, posing a persistent obstacle to precise realization and observation of quantum phenomena.
Controlling errors---or, more broadly, leveraging the information encoded in them for accurate simulation---has therefore been a long-standing yet everlasting subject in quantum science.

{A representative example is digital quantum simulation (DQS)~\cite{suzuki1990fractal,lloyd1996universal,nielsen2010quantum,georgescu2014quantum,altman2021quantum}.}
In DQS, the dynamics generated by a target Hamiltonian composed of generally noncommuting terms, $H{=}\sum_j H_j$, can be approximated through a Trotter decomposition, e.g., $\exp(-i H \delta t) {\approx} \prod_j \exp(-i H_j \delta t)$. A finite step size $\delta t$ inevitably induces Trotter error. Although errors can be suppressed by optimizing the Trotter decomposition~\cite{suzuki1976generalized,suzuki1985decomposition,zhang2023low,eckstein2024large,ikeda2024measuring,zhao2025entanglement,zhang2026taming}, they still accumulate at long times. 
Moreover, realistic quantum devices are affected by additional noise originating from hardware imperfections~\cite{preskill2018quantum}.
Hence, errors induced by both algorithmic approximation and device imperfections eventually drive the system away from the target quantum dynamics. 
Developing noise-robust policies that achieve accurate long-time evolution with limited simulation and measurement resources is thus a central challenge in DQS.

Here, we aim to adopt a perspective that differs from conventional approaches to mitigating Trotter errors. 
Rather than treating deviations from the exact dynamics solely as imperfections to be suppressed, we ask whether these errors---both the intrinsic approximation error of Trotterization and extrinsic sources such as noise---can be brought under constructive control by actively steering the approximate evolution itself. 
In other words, could such errors be harnessed as an efficient resource for their own correction? 
This leads us to the central question: in case this is possible, under which conditions, and with what required resources (measurements, feedback, control operations), can one optimize the entire long-time evolution by adaptively correcting accumulated errors in a resource-efficient way?

Reinforcement learning (RL) provides a natural framework for this task because it is particularly suited to learning policies over long time horizons under stochastic dynamics~\cite{kaelbling1996reinforcement,sutton1998reinforcement,sutton1999policy,lillicrap2015continuous,mnih2015human,arulkumaran2017deep,bukov2026reinforcement}. It has recently been applied to a broad range of important tasks~\cite{chen2013fidelity,fosel2018reinforcement,bukov2018reinforcement,wu2019solving,niu2019universal,zhang2020topological,koch2022quantum,Masahito2020,cao2022quantum,giannelli2022tutorial,sivak2022model,wright2023fast,reuer2023realizing,reinschmidt2023reinforcement,metz2023self,kalita2024domino,cai2025learning,duncan2025taming,sivak2026reinforcement}, including the preparation of spin-squeezed states~\cite{guo2021faster}, the implementation of high-fidelity entangling gates~\cite{baum2021experimental}, and the discovery and experimental realization of previously unknown quantum error-correcting codes~\cite{sivak2023real}.
Such advances demonstrate the ability of RL to optimize long-term objectives in complex quantum-control problems, making it particularly promising for suppressing error accumulation throughout DQS.

Applying RL to achieve adaptive feedback control of DQS nevertheless faces four intertwined challenges. First, a meaningful simulation error must be defined without access to the exact target wave function, which is unavailable precisely in the regime where DQS exceeds classical simulation capabilities~\cite{Bolens2021Reinforcement}. Second, deviations from the target dynamics must be monitored without exponentially costly state tomography~\cite{cramer2010efficient,huang2022learning}. Third, the feedback action space must be expressive enough to correct accumulated errors while remaining sufficiently compact for efficient control~\cite{bukov2018reinforcement,wu2025quantum}. Finally, an effective policy must be trained with limited interactions with the quantum processor, since repeated measurements and control operations introduce substantial experimental overhead~\cite{sivak2022model,khalid2023sample}.

{Here we introduce RL-Trotter to address these challenges and solve} the above-mentioned complex long-time optimization task in DQS (workflow shown in Fig~\ref{fig1_illustration}(a)). In particular, we explicitly demonstrate that low-dimensional physical information from conservation laws, including energy and energy variance, provides a sufficient learning signal to guide an agent. As a result, while approximate evolution inevitably generates errors, RL-Trotter learns to combine them into a self-correcting sequence whose accumulated errors remain controlled over long times. Remarkably, the action space only involves a single scalar, the next Trotter step size, unlike conventional variational quantum algorithms that may
require the simultaneous optimization of hundreds of circuit parameters~\cite{cerezo2021variational,shaffer2023surrogate}. 

RL-Trotter exhibits two properties that are absent in state-of-the-art adaptive algorithms but crucial for experimental implementation: intrinsic robustness against measurement noise and a strong generalization capability.
{The former directly improves training efficiency as it relaxes the precision required for estimating the observables.
Moreover,} the agent only observes a small, fixed set of observables---the energy and energy variance---so its input dimension does not grow with the Hilbert-space dimension. 
{These features substantially reduce the number of measurement shots required for determining a corrective policy, while ensuring that the necessary resources do not scale with system size.}

The generalization capability of RL-Trotter manifests in two distinct ways. Remarkably, a policy trained on a small, classically simulable system can also be transferred directly to systems an order of magnitude larger. Also, a learned policy applies to previously unseen initial states that share similar physical properties with those used in training without further optimization. Such generalization capability thus enables a practical protocol in which RL-Trotter is pretrained classically and then deployed on a large quantum processor, either directly or with limited fine-tuning.
{It actually provides a practical way of implementing an RL strategy on an actual quantum device.}

Our work establishes a new perspective for DQS and quantum algorithms more broadly: unavoidable errors in approximate quantum evolution need not only be suppressed, but can be orchestrated to self-correct. Approximation errors can thereby become resources for realizing accurate long-time quantum dynamics.

\section{Reinforcement-Learning Framework}

Let us outline the RL-Trotter algorithm in detail. Yet, the following discussion can be straightforwardly generalized to different DQS algorithms and model Hamiltonians.

For simplicity, we consider the case where a static Hamiltonian {$H=H_1+H_2$} involves two non-commuting parts,
$H_1$ and $H_2$. By the second-order Suzuki-Trotter formula, the evolution operator over one time step $\delta t$ is given by
\begin{equation}
    \label{eq.Trotter}
    U(\delta t){=}e^{-iH_1\delta t/2}e^{-iH_2\delta t}e^{-iH_1\delta t/2}.
\end{equation} 
{The noncommutativity of $H_1$ and $H_2$ gives rise to Trotter errors, and hence conserved quantities deviate from their initial values.
Our goal is to adaptively select the Trotter step size $\delta t_j$ at each Trotter step, such that DQS can achieve longer simulation times while keeping high long-time simulation accuracy.}

The target dynamics {generated by $H$} feature a set of conserved quantities, e.g., the energy and energy variance of the system
\begin{align}
	\mathcal{E}_n & {=}\langle \psi(t_n) \vert H \vert \psi(t_n) \rangle / L , \notag \\
	\delta \mathcal{E}^2_n &{=}\langle \psi(t_n) \vert H^2 \vert \psi(t_n) \rangle/L {-} L (\mathcal{E}_n)^2, \label{eq:energy_variance}
\end{align}
where $L$ is the system size and $|\psi(t_n) \rangle$ denotes the wavefunction obtained {following} $n$-step Trotter evolution, $\vert\psi(t_n) \rangle {=} \prod_{j=0}^{n-1} U(\delta t_j) \vert\psi(0) \rangle$, with the total time $t_n {=} \sum_{j=0}^{n-1} \delta t_j $. These quantities can be efficiently measured in experiments using, e.g., classical shadows~\cite{huang2020predicting,huang2021efficient}, and do not require access to the exact many-body wave function. 
Indeed, all higher moments of the Hamiltonian should be conserved, but for our purpose the two lowest-order moments are sufficient.

{The existing ADA-Trotter algorithms~\cite{zhao2023making,zhao2024adaptive,zhao2024learning} provide a crucial insight into controlling these errors: maximizing each Trotter step size while bounding deviations in conservation laws, e.g., the energy and energy variance, can yield accurate quantum dynamics. However, this policy is essentially greedy~\cite{mladenov2020optimizing}: a locally (in time) optimal step is, as we demonstrate later, globally suboptimal and exacerbates long-time error accumulation.} We now show explicitly how to go beyond this greedy policy by developing the following RL framework.

We first formulate adaptive Trotterization as a sequential Markov decision process, in which the Trotter step size at each Trotter step is chosen according to the current evolved state.
In general, the choice of the Trotter step size $\delta t_j$ affects not only the error generated at the current step, but also the subsequent quantum state and {future decisions}.
Consequently,
the effect of a single decision propagates throughout the remaining evolution, making the objective, such as the minimization of the Trotter error, {dependent on the long-term consequences of sequential Trotter step sizes rather than on individual local choices}.

Such a process with long-term consequences is precisely
where deep RL excels~\cite{bukov2026reinforcement}.
RL is a subfield of machine learning in which a software agent learns {to solve a task}
by interacting with an environment~\footnote{Here, the term ``environment'' is used in the RL sense, referring to the complete process with which the agent interacts. It should not be confused with the physical environment surrounding the quantum system, such as external degrees of freedom responsible for decoherence and dissipation.} and adapting to its behavior accordingly. The agent generates sequences of actions in the environment and learns to perform {the}
task by maximizing a cumulative reward function {(see Sec.~\ref{sec:ddpg_trotter} of SI for details~\cite{SI})}.

In our setting, the RL environment consists of the quantum system together with the procedure that implements its time evolution, either in a classical simulator or on a quantum processor.
{For simplicity, we train the RL agent using classical simulations in this work. The same framework can, in principle, be extended to in-hardware training on quantum devices, which we discuss later.}

At the $n$th Trotter step, the agent observes the quantum system through expectation values of the conserved quantities and the accumulated evolution time, which define the \textit{observation} $\mathcal{O}_n$ (green arrow in Fig~\ref{fig1_illustration}(a)),
\begin{align}
\mathcal{O}_n =
\left(
\mathcal{E}_n,
\mathcal{E}_0,
\delta \mathcal{E}_n^2,
\delta \mathcal{E}_0^2,
t_n
\right).
\label{eq:rlstate}
\end{align}
Here $\mathcal{E}_n$ and $\delta\mathcal{E}_n^2$ are obtained using the quantum 
state $\lvert\psi(t_n)\rangle$, while $\mathcal{E}_0$ and $\delta\mathcal{E}_0^2$ are their reference values fixed {initially}.
$t_n$ denotes the total evolution time.

The \textit{RL action} $\mathcal{A}_n$ is the time step size $\delta t_n$ (orange arrow in Fig~\ref{fig1_illustration}(a)), chosen within a predefined range $(\delta t_{\min},\delta t_{\max})$ (see Sec.~\ref{sec:MDP_formulation} of SI~\cite{SI}).
Once $\delta t_n$ is determined, the quantum state evolves according to
$
\vert \psi(t_{n+1})\rangle
=
U(\delta t_n)
\vert \psi(t_n)\rangle
$ using the Trotter decomposition in Eq.~\eqref{eq.Trotter} and $
t_{n+1}=t_n+\delta t_n.
$
The agent then receives a \textit{reward} $\mathcal{R}_n$ 
at step $n$ (pink arrow in Fig~\ref{fig1_illustration}(a))
\begin{align}
\mathcal{R}_n {=}
\Bigl[
(1{-}\alpha)
\exp({-}\beta_1\epsilon_{\mathcal{E}_{n+1}}^2)
{+}
\alpha
\exp({-}\beta_2\epsilon_{\delta \mathcal{E}_{n+1}^2}^2)
\Bigr] 
\tanh(\eta \delta t_n),
\label{eq:reward}
\end{align}
with $
\epsilon_{\mathcal{E}_{n}}
    {=}
    \left|\mathcal{E}_n-\mathcal{E}_0\right|,
$ and $
\epsilon_{\delta \mathcal{E}_{n}^2}
    {=}
    \left|\delta\mathcal{E}_n^2-\delta\mathcal{E}_0^2\right|
$
quantifying the error in conserved quantities.
The factor $\tanh(\eta\delta t_n)$ favors larger Trotter step sizes and therefore longer simulated physical times.
For small time steps, $\tanh(\eta\delta t_n) {\simeq} \eta\delta t_n$, so the reward is approximately proportional to the time advanced; for large time steps, it saturates, preventing excessively large steps and stabilizing RL training.
Meanwhile, this reward treats conservation laws as {\it soft constraints} where the exponential factors penalize violations of energy and energy-variance conservation.
The parameter $\alpha$ controls the relative importance of the two constraints, while $\beta_1$ and $\beta_2$ determine the sensitivity to the corresponding conservation-law violations. The parameter $\eta$ controls how strongly the reward favors increasing the Trotter step size~\footnote{The performance of our framework depends on these parameters and also the predefined action range. However, since our reward function is sufficiently simple and informative, we find that the success of training is not sensitive to their specific choice as long as they are chosen within a suitable range.}.

The RL agent aims to learn a policy 
that maximizes the cumulative reward over many Trotter steps. 
Therefore, the agent can choose Trotter step sizes to adaptively correct accumulated errors in energy and energy variance throughout the entire long-time evolution.
We refer to the resulting protocol as \textit{RL-Trotter}.
The learned non-greedy Trotter policy may accept a locally suboptimal choice at a given step to better preserve the conserved quantities over the full evolution.

Before we demonstrate the advantages of RL-Trotter using a concrete model, we first summarize several general features that make RL well suited for DQS.
During training, RL is noise-robust, model-free, and exploration-exploitation balanced. First, RL can optimize the Trotter policy directly from noisy measurement feedback. By training over different noise realizations, the agent learns to maximize the expected cumulative reward and can therefore avoid overreacting to stochastic fluctuations in the measured conservation-law violations. 
Second, as a model-free, feedback-based approach, it can also adapt the Trotter step size without requiring an exact Hamiltonian model, making it applicable when the Hamiltonian parameters cannot be tuned precisely in real experiments. 
{Moreover, RL balances the exploration of different Trotter policies with the exploitation of previously learned experience. It does not require local gradients of the control landscape {with respect to the Trotter step sizes}, which are difficult to estimate on quantum hardware.}
{During evaluation and deployment, the learned policy is generalizable.} As we 
{implement the RL agent using deep neural networks (DNNs)~\cite{huang2021efficient}, RL is able to learn a generalized Trotter policy rather than a protocol tailored to a single quantum evolution.} The same {policy} can therefore be applied to different initial states, quench parameters, noise realizations, and system sizes, reducing the need to redesign or reoptimize the Trotter policy for each simulation.

We use the deep deterministic policy gradient algorithm (DDPG), an off-policy actor-critic~\footnote{The actor network represents the policy that maps the observation $\mathcal{O}_n$ to the next Trotter step size $\delta t_n$, while the critic estimates the expected cumulative reward associated with a state-action pair.}
method, to implement the RL framework for adaptive Trotterization~\cite{silver2014deterministic,lillicrap2015continuous,lillicrap2020continuous}.
The off-policy structure allows trajectories generated during training to be stored in a replay buffer and reused, which improves data efficiency.
This is particularly useful for optimization problems in Trotterization, where the data can be computationally or experimentally costly.
Meanwhile, the dense reward given at each step (cf.~Eq.~\eqref{eq:reward})~\footnote{Although a continuous reward is not strictly required for DDPG, a smoother reward landscape facilitates critic learning and provides more stable policy gradients for the actor.} is more effective for off-policy RL algorithms than a sparse terminal reward, provided only at the final Trotter step~\cite{ng1999policy,andrychowicz2017hindsight}.
Also, DDPG is intrinsically suitable for the continuous action space of $\delta t_n$.
In all cases considered in this work, the policy is learned efficiently, typically within $3000$ training episodes (cf.~Fig.~\ref{fig1_illustration}(b), and Sec~\ref{sec:training_details} of SI~\cite{SI}). Meanwhile, as training progresses {(see videos)}, the agent converges to a stable and near-optimal policy that generates step size sequences enabling longer total evolution time, as shown in Fig.~\ref{fig1_illustration}(c), while keeping violations of the conserved quantities small (cf. Fig~\ref{quench_dynamics}).

In the following sections, we benchmark RL-Trotter against conventional fixed-step Trotterization and ADA-Trotter. We first consider deterministic noise-free simulations, and then study the robustness of the learned policy against measurement noise and its generalization capability to different initial states and system sizes.

\section{Benchmarking RL: mixed-field Ising chain}
\label{Sec:Benchmarking_RL}
For concreteness, we consider a one-dimensional mixed-field Ising model $H=H_1+H_2$ with
\begin{align}
H_1{=}J\sum_{i=1}^{L}\sigma_i^z\sigma_{i+1}^z {+} h_z\sum_{i=1}^{L}\sigma_i^z, \quad 
H_2{=} h_x\sum_{i=1}^{L}\sigma_i^x,
\end{align}
where $\sigma_i^{x,z}$ are Pauli operators and periodic boundary conditions (PBC) are imposed. We set $J{=}1$ as the energy unit.
For small $\vert h_x \vert$ and $\vert h_z \vert$, the antiferromagnetic (AFM) Ising interaction dominates and the ground state lies in the AFM ordered regime. 
Increasing the fields suppresses the staggered order and drives the system to a paramagnetic (PM) regime. 
This model is nonintegrable in the presence of transverse and longitudinal fields, thus serving as a minimal yet nontrivial benchmark for DQS.
Although we focus on this specific model in the following, the {algorithms} 
are general and can be applied to other quantum many-body systems. 

\begin{figure}[t!]
	\centering
	\includegraphics[width=0.49\textwidth]{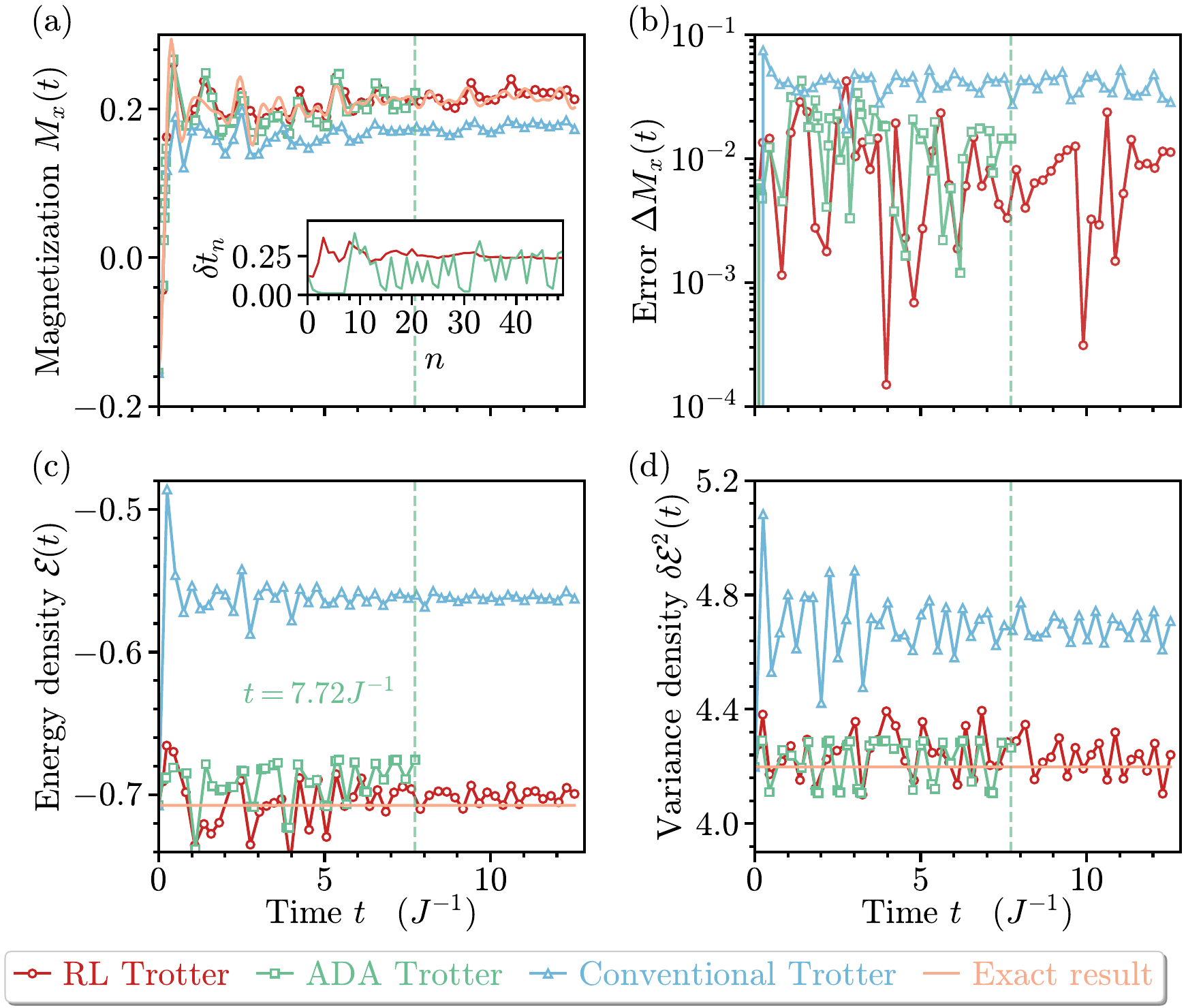}
	\caption{
\textbf{Comparison of the quench dynamics under different Trotter algorithms with exact dynamics (orange).}
(a, b) Dynamics of the magnetization in the $x$ direction and the corresponding error with respect to the exact evolution for the first $50$ Trotter steps. Whereas RL-Trotter (red circles) and conventional Trotter (cyan triangles) have the same total evolution time, $t{=}12.53J^{-1}$,
RL-Trotter produces significantly smaller errors. 
ADA-Trotter (green squares) can suppress the errors to a level comparable to that achieved by RL-Trotter. However, its total evolution time, $t=7.72 J ^{-1}$ shown by the green dashed line, is shorter than that of RL-Trotter as ADA-Trotter freezes, 
i.e., the algorithm tends to always choose the smallest possible step size, 
as shown in the inset of~(a).
(c, d) Dynamics of energy density and energy variance density used as conserved quantities for training.
The average errors in the energy and energy variance density of RL-Trotter are $ \langle \epsilon_\mathcal{E} \rangle \approx 1.19 \times 10^{-2}J$ and $ \langle \epsilon_ {\delta \mathcal{E}^2} \rangle \approx 7.15 \times 10^{-2}J^2$, respectively, which are almost one order of magnitude smaller than those of conventional Trotter, $\langle \epsilon_\mathcal{E}\rangle \approx 1.47 \times 10^{-1}J$ and $ \langle \epsilon_{\delta \mathcal{E}^2} \rangle \approx 4.96 \times 10^{-1}J^2$. 
The tolerances of ADA-Trotter are set to $d_\mathcal{E}=0.03J$ and $d_{\delta \mathcal{E}^2}=0.09J^2$.
RL-Trotter shows a significant advantage over the other two Trotter algorithms. 
The system size is $L=16$.
The quench process is from $h_x = h_z = 0.3$ (AFM) to $h_x = -1.7$ and $h_z = -1.3$ (PM). 
}
\label{quench_dynamics}
\end{figure}

We quench from the AFM ground state of the system with $h_x {=} h_z {=} 0.3 $ to a Hamiltonian with $h_x {=} {-}1.7$ and $h_z {=} {-}1.3$ (PM) for  $N{=}50$ Trotter steps~\footnote{To break the degeneracy of the ground states in the pre-quench Hamiltonian and prepare an initial state with a specific ordering pattern, we add a small perturbation of the staggered field in the $z$ direction to the pre-quench Hamiltonian, i.e., $H_i - \epsilon\sum_i (-1)^i \sigma_i^z$, where $\epsilon$ is a small parameter.}.
Figure~\ref{quench_dynamics} illustrates the dynamics of several representative observables under the RL-Trotter (red). For comparison, we also plot the target exact evolution (orange). As shown in panel (a), 
RL-Trotter generates accurate dynamics of local observables, e.g., the magnetization in the $x$ direction $M_x$: it not only correctly captures the early-time oscillations, but also the relaxation of local observables at
longer times. In panel (b), we plot the error in $M_x$, which is of the order of $10^{-2}$ to $10^{-3}$.

The inset of Fig.~\ref{quench_dynamics}(a) shows the adaptive Trotter step sizes selected by the RL agent. The learned policy automatically chooses smaller step sizes during the initial coherent dynamics and gradually increases them at later times, when the system locally thermalizes and becomes less sensitive to Trotter errors. This highlights the flexibility of the RL framework in adapting the simulation policy to the evolving quantum dynamics.

We attribute the remarkable performance of RL-Trotter to its ability to self-correct violations of the conserved quantities throughout the entire evolution by adaptively adjusting the Trotter step sizes.
In Fig.~\ref{quench_dynamics}(c) and (d), we plot the evolution of energy and energy variance $\mathcal{E}_n$ and $\delta \mathcal{E}_n^2$ and both are well preserved throughout the entire evolution. In particular, both quantities fluctuate around their initial values (orange), indicating that errors accumulated at earlier times can be corrected with adaptive step sizes.
To quantify the averaged error over the entire evolution, we define the average error $\langle \epsilon \rangle {=} \sum_{n=1}^N \epsilon_n /N$ and find
$\langle \epsilon_\mathcal{E} \rangle{\approx} 1.19{\times} 10 ^{-2} J $ and $\langle \epsilon_{ \delta \mathcal{E}^2} \rangle {\approx} 7.15{\times} 10 ^{-2} J^2 $. 

We further find that accurately constraining the energy and energy variance automatically suppresses violations of higher-order energy moments (not shown). Heuristically, this can be understood from the central-limit theorem of large many-body systems~\cite{CLT2004Hartmann,zhao2023making}: the energy distribution becomes increasingly well characterized by its first two moments, while higher-order cumulants provide only subleading corrections. Consequently, the energy and energy variance capture the dominant information governing the dynamics, allowing RL-Trotter to construct a faithful approximation to the target generator and thereby accurately reproduce the long-time evolution of local observables.

To show the advantage of the RL-Trotter over the conventional Trotter,
we simulate the dynamics using the fixed-step Trotter {(cyan triangles in Fig.~\ref{quench_dynamics})} with the same total evolution time $t{=}12.53J^{-1}$ and total number of Trotter steps. 
Unlike RL-Trotter, it cannot adaptively correct errors once the step size $\delta t {=} t_N/N$ is fixed.
{
The resulting averaged errors read
$\langle \epsilon_\mathcal{E} \rangle {\approx} 1.47\times10^{-1}J$
and
$\langle \epsilon_{\delta \mathcal{E}^2} \rangle {\approx} 4.96 \times10^{-1}J^2$, and $\langle \epsilon_{M_x}\rangle$ is on the order of $10^{-2}$ to $10^{-1}$. All these errors are approximately one order of magnitude larger than their RL-Trotter counterparts.
}

Finally, we also compare the RL-Trotter with ADA-Trotter (green squares in Fig.~\ref{quench_dynamics}). In short, ADA-Trotter greedily maximizes the step size $\delta t_n$ at each Trotter step via a feedback loop, while treating the conservation laws of energy and energy variance as {\it hard constraints}, using some predefined fixed tolerances $d_\mathcal{E}$ and $d_{\delta\mathcal{E}^2}$, respectively. To make a fair comparison, we choose $d_\mathcal{E}{=}0.03J$ and $d_{\delta \mathcal{E}^2}{=}0.09J^2$, such that ADA-Trotter suppresses the error to the same level as RL-Trotter, e.g., $\langle \epsilon_\mathcal{E} \rangle {\approx} 1.93 {\times} 10^{-2} J $ and {$\langle \epsilon_{M_x} \rangle {\approx} 1.41 {\times} 10^{-2}$.} However, due to the strong constraints on conserved quantities, ADA-Trotter often ‘freezes’, i.e., the algorithm tends to always choose the smallest possible step size (see the inset of Fig.~\ref{quench_dynamics}(a)). Consequently, the quantum state barely propagates, despite the large number of quantum gates consumed.
ADA-Trotter reaches the total evolution time $t {=} 7.72 J^{-1}$, which is only around {$60\%$} of that of RL-Trotter.

Freezing originates from the locally greedy nature of ADA-Trotter. A locally (in time) optimal step is likely globally suboptimal
and exacerbates long-time error accumulation. In contrast, RL-Trotter may temporarily tolerate small deviations in the conserved quantities (see the first two red circles in Fig.~\ref{quench_dynamics}(c)) and subsequently correct them, such that the cumulative reward over the entire trajectory is optimized. This global optimization policy enables RL-Trotter to achieve longer evolution times while maintaining accurate control of the conserved quantities. A comparison with ADA-Trotter using different error tolerances is provided in Sec.~\ref{subsec:ADA_with_different_tolerance} of the SI~\cite{SI}, where we show that achieving comparable evolution times with ADA-Trotter requires relaxing the tolerances, leading to larger deviations in the conserved quantities.

\begin{figure}[t!]
	\centering
	\includegraphics[width=0.49\textwidth]{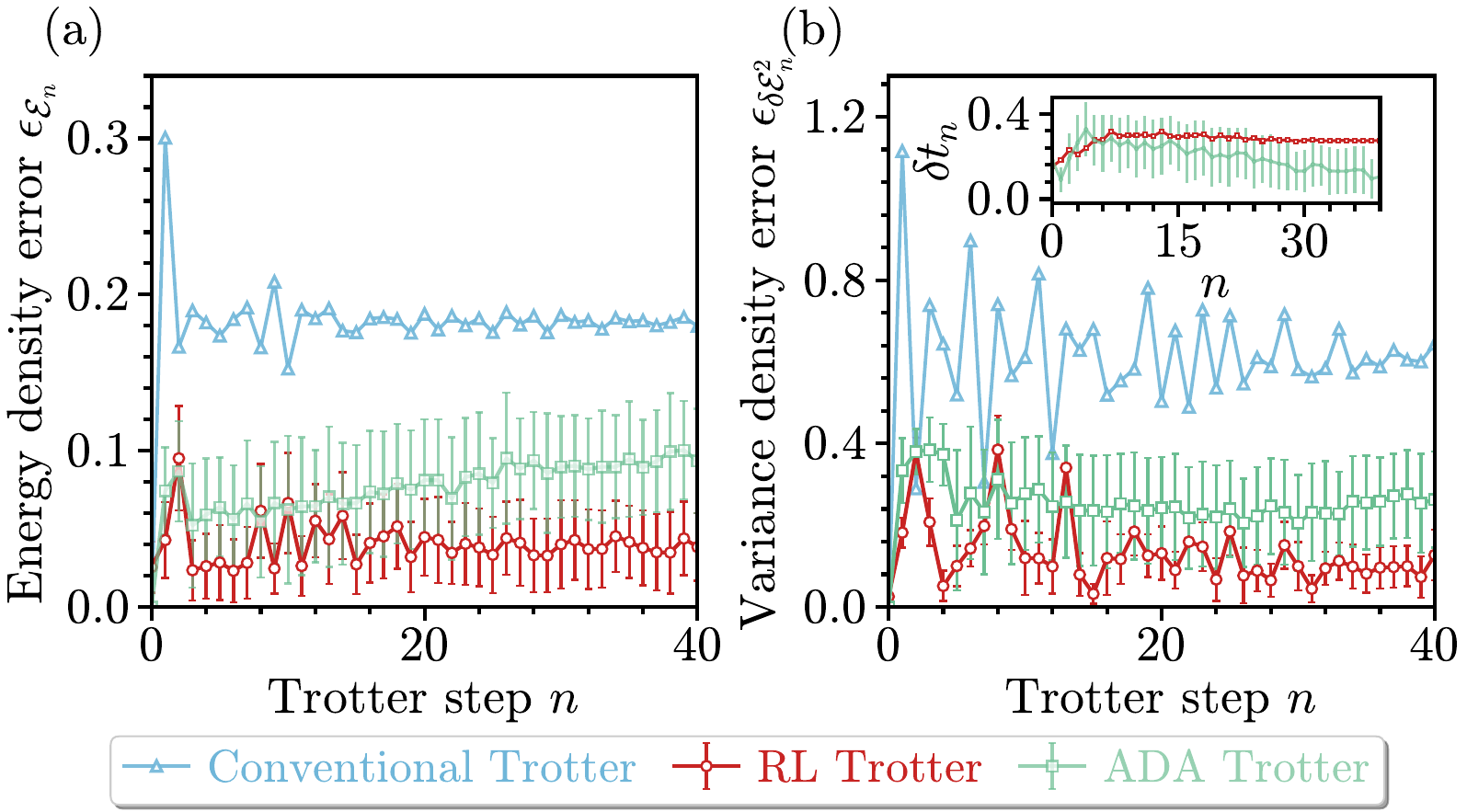}
	\caption{
\textbf{Noise robustness of different Trotter algorithms.} 
We plot the errors in the energy density (a) and energy variance density (b) during the first $40$ Trotter steps for different Trotter algorithms in a noisy environment with noise strength $\sigma_1 {=}\sigma_2 {=}0.03$. 
The RL agent is trained in a noisy environment with $\sigma_1 {=}\sigma_2 {=}0.1$. 
For RL-Trotter (red) and ADA-Trotter (green), the markers and error-bar regions denote the mean and standard deviation over $50$ independent evaluations. 
The inset of (b) shows the Trotter step size $\delta t_n$ at each Trotter step. 
RL-Trotter remains stable against noise throughout the evolution, whereas the average Trotter step size of ADA-Trotter decreases continuously. 
The total evolution time of RL-Trotter is $t {=} 11.04 \pm 0.09 J^{-1}$, approximately $ 40 \% $ longer than that of ADA-Trotter, $t {=} 7.94 \pm 1.42 J^{-1}$. 
At the $40$th Trotter step, the errors of RL-Trotter are $\epsilon_\mathcal{E} {=} (3.84 \pm 2.16){\times}10^{-2}J$ and $\epsilon_{\delta \mathcal{E}^{2}} {=} (1.27 \pm 0.62){\times}10^{-1}J^2$, while those of ADA-Trotter are $\epsilon_\mathcal{E} {=} (9.34 \pm 3.34){\times}10^{-2}J$ and $\epsilon_{\delta \mathcal{E}^{2}} {=} (2.63 \pm 1.18){\times}10^{-1}J^2$.
As a reference, we also consider conventional Trotter with the same average total evolution time as RL-Trotter, $t {=} 11.04 J^{-1}$ in a deterministic environment.
Its errors are $1.80{\times}10^{-1}J$ and $6.44{\times}10^{-1}J^2$ for the energy density and energy variance density, respectively. The system size is $L=16$.
    }
\label{Robustness}
\end{figure}
\begin{figure}[t!]
	\centering
	\includegraphics[width=0.49\textwidth]{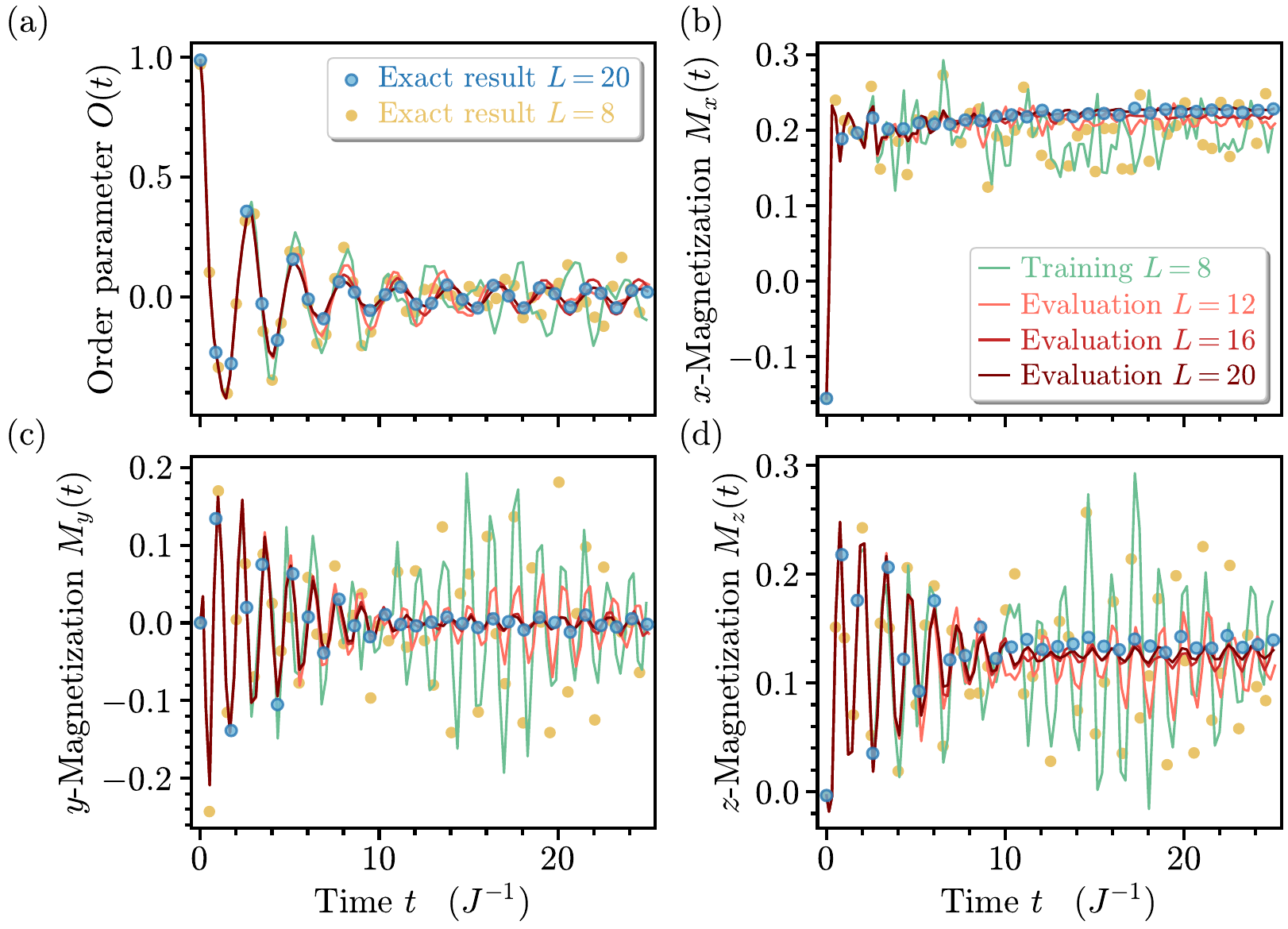}
	\caption{
    \textbf{Generalization of the learned policy to larger system sizes.}
We plot RL-Trotter dynamics of (a) the order parameter $O(t)$ and (b--d) the magnetization $M_{\alpha}(t)$ with $\alpha{=}x,y,z$ for different system sizes. 
The policy is trained only at $L{=}8$, whose results are shown by the green curves. 
We then apply the learned policy, without further training, to larger systems with $L{=}12,16,20$ (red curves). 
The exact time-evolution curves for $L{=}8$ and $L{=}20$ are shown by the yellow and blue dots, respectively. 
Although the thermalization behavior depends strongly on the system size, the RL-Trotter dynamics for $L=20$ agree well with the exact dynamics at the same system size, demonstrating the transferability of the learned policy to larger systems.
We consider the quench from $h_x{=}h_z{=}0.3$, to $h_x{=}{-}1.7$, $h_z{=}{-}1.3$ over the first $100$ Trotter steps.
	}
\label{scaling_effect}
\end{figure}

\section{Robustness and Generalization}
Having established the effectiveness of RL-Trotter for a concrete noise-free system, we now examine its robustness and generalization capabilities, which are essential for practical implementation on quantum devices. 

In general, applying optimization-based control to quantum hardware faces two major challenges. First, quantum devices are inherently noisy, and measurement errors can distort the physical observables used to guide adaptive optimization, leading to unreliable feedback control. Second, optimization protocols often require repeated training for different system sizes, Hamiltonian parameters, or initial conditions, resulting in substantial computational and experimental overhead. It therefore remains an open question whether the resources devoted to such case-by-case optimization can be reduced by learning transferable control policies.
 
RL-Trotter addresses both challenges. The learned policy remains robust against measurement noise and exhibits remarkable generalization across system sizes and initial states, substantially reducing the resources required for repeated optimization; {these are inherent features of learning algorithms, such as RL, as compared to conventional optimization.} Although our demonstrations are based on classical emulation, these features suggest a promising route toward feedback-based Trotter control on real quantum devices.

\subsection{Robustness against measurement noise}
\label{sec:Robustness}

RL is particularly well suited to deal with measurement noise, which can arise from the finite number of  measurements used to estimate the conserved quantities, for two reasons.
First, RL optimizes the expected cumulative reward rather than the reward obtained in a single episode~\cite{bukov2026reinforcement}, so stochastic fluctuations can  be effectively averaged out during training. 
Second, noise can promote exploration and help prevent the agent from becoming trapped in overly conservative policies.

We examine the robustness of the learned policy in noisy environments from two complementary perspectives.
First, we test whether an agent trained in a noisy environment can learn an effective policy. 
Second, we test whether the learned policy remains robust under noise realizations different from those encountered during training.

To mimic the measurement noise {while remaining agnostic to the specifics of a concrete quantum device type}, we add Gaussian noise to the measured constrained quantities in classical emulation, $\mathcal{E}_n {\rightarrow} \mathcal{E}_n +\xi_1$ and $\delta \mathcal{E}^2_n {\rightarrow} \delta \mathcal{E}^2_n+\xi_2$,
where $\xi_1 {\sim} \mathcal{N}(0,\sigma_1^2)J$ and $\xi_2 {\sim} \mathcal{N}(0,\sigma_2^2)J^2$ are independent Gaussian random variables with zero mean and standard deviations $\sigma_1$ and $\sigma_2$, respectively.
In particular, we train the agent with relatively strong noise, $ \sigma_1 {=}\sigma_2 {=}0.1$,
and evaluate the learned policy using a weaker but finite noise strength, $\sigma_1 {=}\sigma_2 {=}0.03$. 
In practice, the noise strength can be estimated given a finite measurement budget. For instance, the noise level $ \sigma_1 {=}\sigma_2 {=}0.1$ approximately corresponds to $10^2$ and $10^4$ measurement shots for energy and energy variance, respectively, using random measurement techniques, e.g., classical shadows~\cite{huang2021efficient}. 

We repeat the noisy evolution $50$ times and plot the mean and standard deviation of the Trotter error at each Trotter step for RL-Trotter (cf. Fig.~\ref{Robustness}, red dots; each trajectory is shown in Sec~\ref{sec:Robustness_comparison} of SI~\cite{SI}).
After the initial oscillatory regime, 
the mean Trotter errors $\epsilon_{\mathcal{E}_n}$ and $\epsilon_{\delta\mathcal{E}^2_n}$ approach stable values of approximately $\langle \epsilon_{ \mathcal{E}} \rangle {\approx} 0.04J$ and $ \langle  \epsilon_{ \delta \mathcal{E}^2} \rangle {\approx} 0.13J^2$, respectively.
These errors are slightly larger than those in the noise-free case.
Because the measured observables are affected by measurement noise, the observation received by the RL agent varies between different evaluation runs. The agent therefore selects slightly different Trotter step sizes in different runs, leading to small run-to-run fluctuations in the total evolution time.
The resulting total evolution time is $t{=}11.04{\pm} 0.09J^{-1}$, where the uncertainty denotes the standard deviation over $50$ independent evaluations. 

For comparison, the conventional Trotter (cyan triangles) with the same total evolution time $t{=}11.04J^{-1}$ generates much larger errors $\langle\epsilon_{ \mathcal{E}} \rangle {\approx} 0.18J$ and $\langle
\epsilon_{ \delta\mathcal{E}^2 } \rangle {\approx} 0.61J^2$. 
{In addition, RL-Trotter substantially outperforms ADA-Trotter in the presence of measurement noise.}
For ADA-Trotter, we set fixed tolerances to $d_{\mathcal{E}}{=}0.08J$ and $d_{\delta\mathcal{E}^2}{=}0.4J^2$, and repeat the noisy evolution $50$ times {(see Sec.~\ref{sec:Robustness_comparison} of the SI for details~\cite{SI})}.
These two constraints are much looser than the stable error levels achieved by RL-Trotter.
They yield mean ADA-Trotter errors of $\langle\epsilon_\mathcal{ E} \rangle {\approx} 0.08J$ and $\langle \epsilon_{ \delta \mathcal{E}^2} \rangle {\approx} 0.25J^2$, nearly twice as large as those of RL-Trotter.
Nevertheless, the total evolution time of ADA-Trotter is $t=7.94{\pm}1.42 J^{-1}$, where the mean value is much shorter than that of RL-Trotter, albeit with much larger run-to-run-fluctuation strength.

{The superior performance of RL-Trotter originates from its robustness against noisy estimates of the conserved quantities.
In ADA-Trotter, measurement noise in the energy and energy variance directly affects the adaptive selection of the Trotter step, leading to inaccurate control of the conservation laws. 
Consequently, the chosen Trotter step size exhibits large fluctuations and sometimes freezes in a single realization; 
the averaged step sizes also  progressively decrease. 
In contrast, RL-Trotter learns a global optimization policy that is substantially less sensitive to measurement noise, resulting in both longer accessible simulation times and improved control of Trotter errors.}

\subsection{Generalization to larger system sizes}

We now train the RL agent on a small system with $L{=}8$ spins, and show that the learned policy can be directly applied to larger systems with $L{=}12,16,20$ without further training.
We plot the expectation value of the order parameter ${O}(t) = \langle \psi(t) \vert\sum_i (-1)^i \sigma_i^z\vert\psi(t) \rangle/L$, which quantifies the memory of the AFM order,
and the magnetization $M_{\alpha}(t)$ with $\alpha{=}x,y,z$ for different system sizes in Fig.~\ref{scaling_effect}. For $L{=}8$ (green curves), at early times local observables relax quickly, while persistent temporal fluctuations survive even at long times due to finite-size effects. We have also confirmed that these local dynamics closely follow the exact solution (yellow dots). 
In contrast, for larger systems, such fluctuations damp out in time as the system locally thermalizes, with observables approaching their thermal equilibrium prediction; see the exact simulation results (blue dots, $L{=}20$) in Fig.~\ref{scaling_effect}.

Importantly, although the local evolution exhibits a strong dependence on the system size, the agent trained on a small system is directly applicable to larger systems; see red curves in Fig.~\ref{scaling_effect}, where darker colors represent larger system sizes. In particular, the results for $L{=}20$ (dark red) follow the exact results (blue) closely throughout the entire evolution. {In contrast, since the feedback loop in ADA-Trotter determines a specific sequence of Trotter step sizes for a given system size rather than a policy, the resulting sequence is no longer applicable when transferred to larger systems; see Sec.~\ref{subsec:failure_generalization_ADA} of SI~\cite{SI} for details.}

As a second example, we consider a quench process where the post- and pre-quench Hamiltonians are within the same AFM phase. 
The corresponding entanglement entropy grows relatively slowly such that we are able to numerically simulate the dynamics involving $L{=}100$ sites using matrix product state (MPS) techniques~\cite{schollwock2011density}.
{We provide the main results here and illustrate the details of MPS in Sec.~\ref{sec:Generalization_to_large_systems_via_MPS} of SI~\cite{SI}.}

The results are shown as red dots in Fig.~\ref{MPS_main}. Remarkably, RL-Trotter (red) can still self-correct accumulated errors and accurately reproduce the exact dynamics (orange curves) in both conserved quantities and local observables, even though now the system size is {an order} of
magnitude larger than the one used for training the RL agent.
For comparison, we also show the results of the fixed-step Trotter (cyan triangles) using the same total evolution time and the same number of Trotter steps. 
RL-Trotter reduces errors in different observables by approximately one order of magnitude compared with the conventional Trotter scheme.

\begin{figure*}[t!]
	\centering
	\includegraphics[width=0.8\textwidth]{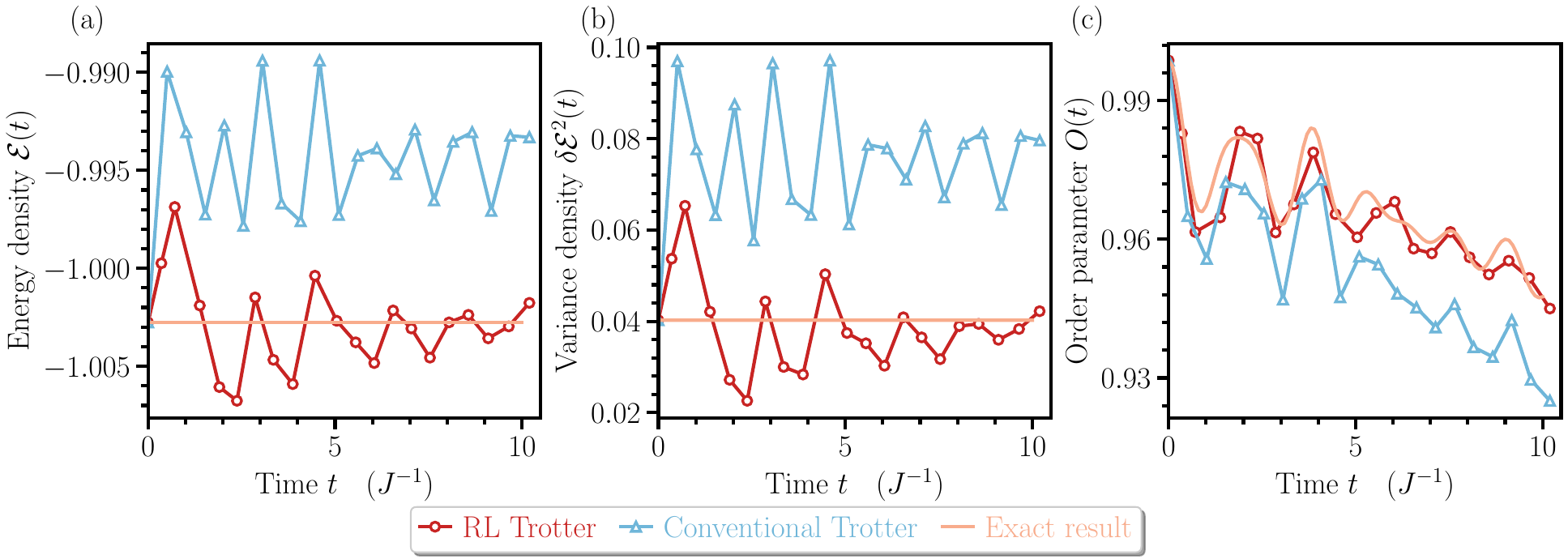}
	\caption{\textbf{MPS benchmark of the RL-Trotter protocol for a large system.} Time evolution of (a) the energy density $\mathcal{E}(t)$, (b) the energy-variance density $\delta \mathcal{E}^2(t)$, and (c) the staggered order parameter $O(t)$ for a system with $L=100$ over the first $20$ Trotter steps.
The RL agent is trained on small systems of $L=8$ using ED with periodic boundary conditions. We directly apply this agent to a large system using MPS representation with open boundary conditions.
The quench is performed from $h_z=0$ and $h_x=0.1$ to $h_z=0.3$ and $h_x=0.3$. 
The average errors of $\mathcal{E}$, $\delta \mathcal{E}^2$, and the AFM order $O$ are $1.62\times10^{-3}J$, $7.07\times 10^{-3}J^{2}$, and $3.47\times10^{-3}$, respectively. All error measures are reduced by approximately one order of magnitude compared with those of conventional Trotter, which are $8.15 \times 10^{-3}J$, $3.46\times10^{-2}J^2$, and $1.39 \times 10^{-2}$, {respectively}. 
{These results demonstrate that the RL agent trained on small systems can accurately optimize DQS of systems {an order} of magnitude larger.}	
}
\label{MPS_main}
\end{figure*}

\begin{figure}[t!]
	\centering
	\includegraphics[width=0.48\textwidth]{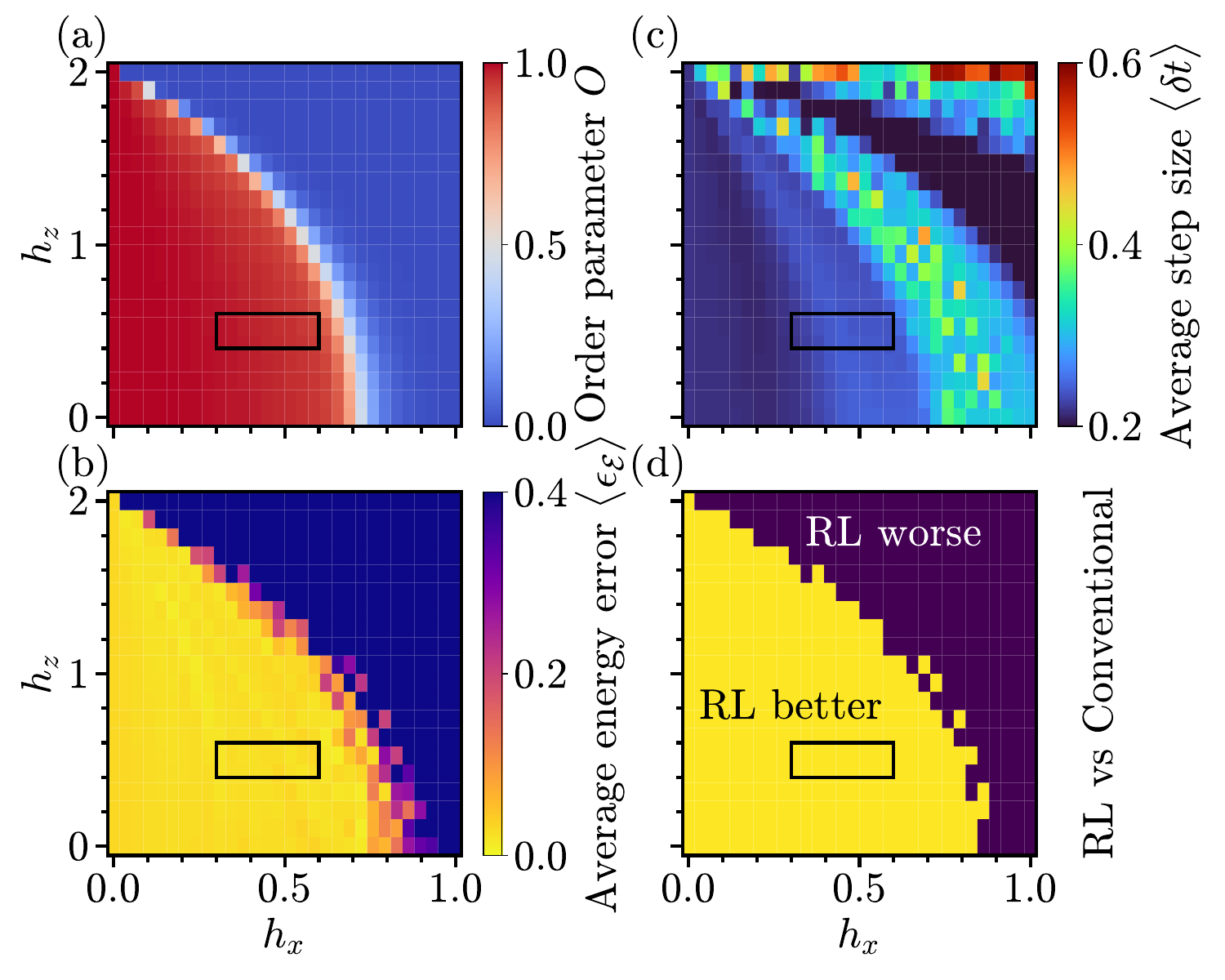}
	\caption{
\textbf{Generalization of the learned policy across different Hamiltonian parameters.} 
The policy is trained on quenches from initial ground states randomly sampled in the AFM region $0.3J{<}h_x{<}0.6J$ and $0.4J{<}h_z{<}0.6J$ (black box) to a fixed postquench point, $h_x{=}{-}1.7J$ and $h_z{=}{-}1.3J$, deep in the PM phase {(outside axes boundaries)}.
We then apply the learned policy, without further training, to a broader region, $0{<}h_x{<}1$ and $0{<}h_z{<}2$ (the entire plot boundaries), and evaluate the evolution behavior over the first $40$ Trotter steps. 
{(a) Initial-state staggered order parameter $O$,} (b) Average Trotter error of the energy density, $\langle \epsilon_\mathcal{E} \rangle$, and (c) average time step, $\langle \delta t \rangle$, as functions of the initial Hamiltonian parameters $h_x$ and $h_z$. 
The successful generalization region, where the Trotter errors remain small while the average Trotter step size remains sufficiently large to generate time evolution, coincides with the region of strong antiferromagnetic order, as indicated by $O$ in (a).
Specifically, the average Trotter step size in this region satisfies $\langle \delta t\rangle {\in} [0.21,0.36]J^{-1}$, as shown in (c).
In (d), we compare the average Trotter error of the energy density, $\langle \epsilon_\mathcal{E} \rangle$, with that of conventional Trotter using the same total evolution time as RL-Trotter. 
The yellow region indicates where RL-Trotter has a smaller $\langle \epsilon_\mathcal{E} \rangle$ than conventional Trotterization. 
We find that the better performance of RL-Trotter is strongly correlated with the staggered order parameter of the initial state, which suggests that the learned policy transfers to initial states with similar AFM order, rather than being tied to a single set of Hamiltonian parameters.
The system size is $L=18$.
}
\label{generalization}
\end{figure}

\subsection{Generalization to different parameter settings}
So far, we have always considered a fixed initial state and post-quench Hamiltonian. In practice, however, one may wish to simulate dynamics starting from a broad family of initial states, e.g., states belonging to the same phase. Training a separate agent for each specific setting would incur substantial computational costs. We now show that RL-Trotter can learn a physically transferable policy that applies directly to different initial states that are physically related, including states not encountered during training.

To demonstrate this, we now aim to study the quench dynamics where the pre-quench Hamiltonian is randomly selected from the parameter region $0{<}h_x{<}1$ and $0{<}h_z{<}2$. The corresponding ground state will be used as the initial state, which spans both AFM and PM regimes; see Fig.~\ref{generalization}(a) where we plot the AFM order parameter $O$ of the initial states. The post-quench Hamiltonian is still fixed at $h_x{=}-1.7J$ and $h_z{=}-1.3J$, deep in the PM phase.

During training, to improve the generalization capability of the RL agent, we randomly sample a pre-quench Hamiltonian within a narrow parameter region in the AFM phase ($0.3J{<}h_x{<}0.6J$ and $0.4J{<}h_z{<}0.6J$, the black box in Fig.~\ref{generalization}), and use the corresponding ground state as the initial state for different training episodes.
The RL agent is then trained to maximize the average reward over this ensemble of initial states. 
Note that now the instantaneous values of $\mathcal{E}_n$, $\delta \mathcal{E}_n^2$, and $t_n$ do not necessarily determine a unique optimal Trotter step, because different initial states may exhibit different dynamics even when these quantities coincide. We therefore include the initial values of the conserved quantities in the observation, cf.~Eq.~\eqref{eq:rlstate}. This additional information allows a single policy to distinguish among different initial conditions and select an appropriate Trotter policy for each of them.

After training, we evaluate the learned policy throughout the entire parameter region $0{<}h_x{<}1$, and $0{<}h_z{<}2$. Figure~\ref{generalization}(b) shows the average Trotter error in the energy density, $\langle\epsilon_\mathcal{E}\rangle$, over $40$ Trotter steps. Comparison with the staggered order parameter in Fig.~\ref{generalization}(a) reveals that the error remains small throughout the AFM region. This observation thus suggests
that the agent learns a policy transferable among initial states beyond the ones included in the training set, as long as they share the same underlying physical {properties}. 
This conclusion is further supported by Fig.~\ref{generalization}(d), which compares the averaged energy errors induced by RL-Trotter and conventional Trotter under matched simulation resources. The yellow region, where RL-Trotter produces a smaller energy error, largely coincides with the parameter region supporting the AFM phase. 

Figure~\ref{generalization}(c) provides a complementary view through the average Trotter step size, $\langle\delta t\rangle$. 
Within the AFM region, $\langle\delta t\rangle$ varies only weakly across the parameter space, showing that the learned policy selects a consistent adaptive policy while maintaining small Trotter errors. In the PM region, by contrast, the selected Trotter step sizes fluctuate strongly with the pre-quench parameters and are accompanied by substantially larger errors. The policy therefore does not transfer reliably to initial states whose physical character differs fundamentally from that of the training states. These results also emphasize that a large Trotter step size alone does not indicate successful generalization; it must be accompanied by accurate control of the conserved quantities.

\section{discussion and outlook}
In this work, we introduce self-correcting RL-Trotter, a reinforcement learning framework to realize faithful long-time simulation of DQS. Its effectiveness rests on two central ideas.
{First, the unavoidable Trotter approximation errors, which are characterized by the deviations of the energy and energy variance from their initial values, provide a compact representation of the simulation information and sufficient learning signals to guide the agent.
This information efficiency is consistent with the central-limit theorem of many-body systems~\cite{CLT2004Hartmann}, for which the energy distribution can often be characterized predominantly by its lowest two moments.}
Second, RL optimizes the cumulative reward over the full trajectory rather than enforcing the constraints independently at each step. 
It can therefore coordinate decisions across the evolution, and thereby achieve better long-time control by self-correcting accumulated Trotter errors.

{The experimental implementation of feedback-control algorithms is often constrained by the limited measurement resources available on quantum devices. RL-Trotter is particularly suitable in this regard, as the agent does not require access to the complex many-body wavefunction but only the energy and energy variance, which can be efficiently estimated using randomized measurement techniques such as classical shadows. Even with finite measurement budgets, such as $10^2$ shots for the energy and $10^4$
shots for the energy variance at each feedback step considered in Sec.~\ref{sec:Robustness}, the learned policy remains effective and substantially outperforms other Trotterization algorithms. 
Moreover, in a realistic quantum device, additional imperfections, including uncertainties in Hamiltonian parameters, gate errors, and calibration drift, will also affect the feedback process, yet have not been included in the current demonstration. Nevertheless, the observation-based and model-free formulation of RL-Trotter provides a natural route to incorporate such device-specific imperfections directly into the training process and develop tailored RL-Trotter protocols with improved performance.}

The generalization capability of RL-Trotter further reduces the experimental overhead of optimization.
In particular, the fact that a policy learned from small systems can be directly used to control systems an order of magnitude larger is truly remarkable.
Understanding its origin and extending it to broader classes of Hamiltonians and initial states represent important directions for future work. 
This also suggests  
a practical experimental policy: one first pretrains the agent using classical simulations, while incorporating available information about the target hardware, such as measurement noise, 
into the simulated training environment; 
the resulting policy can then be transferred directly to the quantum device of a large size. 
When additional optimization is required for a particular system or parameter regime, the original policy can serve as an effective initialization for on-device fine-tuning, which is expected to require substantially fewer experimental episodes and measurements than training from scratch. 
In this way, the robustness and generalization capability of RL-Trotter together provide a resource-efficient route toward self-correcting digital quantum simulation on realistic quantum hardware.

{Previous studies have highlighted the crucial role of the Hamiltonian~\cite{childs2021theory}, initial states~\cite{csahinouglu2021hamiltonian,mizuta2025trotterization,chen2024average,mobus2024strong,su2021nearly}, and quantum resources~\cite{zhao2025entanglement,zhang2026taming} in determining quantum simulation errors.
In contrast, 
our results demonstrate that the accurate control of quantum simulation should continuously account for the low-dimensional errors generated throughout the Trotterized evolution, instead of focusing solely on the information available at the initial stage.}
Several extensions within the present framework are worth investigating. Beyond the energy and its variance, additional conserved quantities may provide complementary information about simulation errors. 
Possibilities include globally conserved charges associated with symmetries {as well as} local constraints, such as Gauss-law conditions in lattice gauge theories~\cite{zohar2016quantum,aidelsburger2021cold,fragomeno2026implementing}. 
Incorporating these additional diagnostics into the observations and reward could enable a more accurate RL-Trotter algorithm. 
In addition, the action space can be expanded beyond the Trotter step size: the agent could adaptively select different DQS algorithms, e.g., higher-order Trotter decompositions~\cite{suzuki1990fractal}
and random compilers~\cite{nakaji2024high}, as the dynamics evolve.
Generalization of RL-Trotter to time-dependent Hamiltonians should be straightforward. Although the energy is generally not conserved under explicitly time-dependent driving, one can construct piecewise conservation laws to guide the RL optimization process~\cite{zhao2024adaptive}.

More broadly, extending self-correcting RL beyond DQS remains an important direction. 
In other quantum algorithms, conserved quantities may be replaced or supplemented by problem-specific physical or algorithmic constraints, whose violations can define the observations and rewards guiding learning. 
This framework could apply, for example, to variational quantum eigensolvers~\cite{peruzzo2014variational} or constrained quantum-search algorithms~\cite{gilliam2021grover}.
An important open question is whether such limited constraint information is sufficient to optimize quantum algorithms when exact solutions are unavailable.

\textit{Acknowledgments.---}%
We thank S. L. Ren
for estimating measurement costs. We also thank Y. X. Zhang, C. H. Lee, and Pan Zhang for helpful discussions.
This work is supported 
by Quantum Science and Technology-National Science and Technology Major Project
(No. 2024ZD0301800) and
by the National Natural Science
Foundation of China (Grant No. 12474214), and by High-performance Computing Platform of Peking University. Part of the numerical simulations were performed on the MPIPKS HPC cluster.
MB was funded by the European Union (ERC, QuSimCtrl, 101113633). Views and opinions expressed are, however, those of the authors only and do not necessarily reflect those of the European Union or the European Research Council Executive Agency. Neither the European Union nor the granting authority can be held responsible for them. 
This work was in part supported by the Deutsche Forschungsgemeinschaft under grants
FOR 5919 (project-id 550495627), and the cluster of excellence ct.qmat (EXC 2147, project-id 390858490).
The authors used ChatGPT 5.6 and ChatGPT 6.0 Astra during the preparation of this manuscript to improve readability and assist with code optimization.
The authors assume full responsibility for the scientific content, accuracy, and integrity of this paper.

\textit{Contributions.---}%
YBS designed the RL framework and performed the numerical simulations. 
HZ and MB supervised this work. 
All authors contributed to discussions of the results, the development of the project, and the drafting of the manuscript.

\textit{Data and Code Availability.---}%
The data, code, and trained models associated with this manuscript are deposited in Zenodo (DOI: \href{https://doi.org/10.5281/zenodo.22037440}{10.5281/zenodo.22037440}).
Automatic access is restricted to prevent automated data and code harvesting by artificial intelligence systems. 
Access credentials will be provided to human researchers upon request via Zenodo for academic and verification purposes.

\ifcombinedmode
\bibliography{bibfile}
\else
\maketitle
\fi 

\else 
\fi 

\resumetoc
\ifsuppmode

\cleardoublepage

\ifcombinedmode

\makeatletter
\let\maketitle\savedmaketitle
\let\author\savedauthor
\let\affiliation\savedaffiliation
\let\email\savedemail
\let\thanks\savedthanks
\makeatother
\else

\fi 

\setcounter{affil}{0}
\renewcommand{\thefigure}{S\arabic{figure}}
\setcounter{figure}{0}
\renewcommand{\theequation}{S.\arabic{equation}}
\setcounter{equation}{0}
\renewcommand{\thesection}{S\arabic{section}}
\setcounter{section}{0}

\onecolumngrid
\begin{center}
\textbf{\large{\textit{\nqdcolor{Supplementary Material}}}} \\ 
\end{center}

\twocolumngrid

\ifcombinedmode
\title{Reinforcement Learning to Harness Approximation Errors for Long-Time Quantum Simulation}
\author{Yu-Bo Shi}
 \affiliation{\affA}
  \affiliation{\affD}
	\author{Markus Heyl}
 \affiliation{\affC}
	\author{Roderich Moessner}
 \affiliation{\affB}
  \author{Marin Bukov}
   \email{mgbukov@pks.mpg.de}
 \affiliation{\affB}
\author{Hongzheng Zhao}
\email{hzhao@pku.edu.cn}
 \affiliation{\affA}
 	\date{\today}

\maketitle
\else
\maketitle
\fi 

\begin{widetext}

\tableofcontents

\section{Supplemental information on the deep deterministic policy gradient algorithm for RL-Trotter}
\label{sec:ddpg_trotter}
 
In this section, we provide supplementary details of the deep deterministic policy gradient (DDPG) algorithm~\cite{lillicrap2020continuous,silver2014deterministic} used in our work. DDPG is an off-policy actor-critic algorithm designed for continuous action spaces, making it well suited to RL-Trotter, in which the Trotter step size at each step is a continuous variable.
{DDPG employs two deep neural networks: an actor, which maps the current observation to an action, and a critic, which estimates the expected return associated with the observation–action pair.}
We first briefly formulate the adaptive Trotterization problem as a Markov decision process (MDP) and then describe the structure and optimization procedure of DDPG in detail. We further specify the training details, and explain the motivation behind the reward design.

\subsection{MDP formulation of RL-Trotter}
\label{sec:MDP_formulation}

{We formulate adaptive Trotterization as an MDP, in which the Trotter step size at each step is determined by the RL agent.}
At the $n$th Trotter step, the agent first gets the observation $\mathcal{O}_n$ (cf. Eq.~\eqref{seq:RL_observation} below),
constructed from the current quantum state.
Based on this observation, the agent selects an RL action $\mathcal{A}_n$, which is the Trotter step size $\delta t_n$ at the $n$th Trotter step.
After the action is chosen, the quantum state is evolved according to the corresponding Trotter decomposition. We then obtain the next observation $\mathcal{O}_{n+1}$.
After that, a reward $\mathcal{R}_n$ is returned to the agent based on the previous observation, the chosen action, and the resulting observation, thereby evaluating the quality of the current action. 
The agent then observes the updated state and selects the next action. This process is repeated until all $N$ Trotter steps are completed.
{The entire sequence is called an episode. }
The objective of the agent is to maximize the expected cumulative RL reward over the whole episode,
rather than to optimize only the immediate reward at each individual step. 
We now describe this process in detail and specify the roles of the actor and critic networks of DDPG.

Based on the quantum state at the $n$th step $\vert \psi_n(t_n) \rangle$, we can measure the observation
\begin{align}
\label{seq:RL_observation}
\mathcal{O}_n =
\left(
\mathcal{E}_n,
\mathcal{E}_0,
\delta \mathcal{E}_n^2,
\delta \mathcal{E}_0^2,
t_n
\right).
\end{align}
We feed the observation $\mathcal{O}_n$ into the actor network, which outputs a deterministic value $\mu(\mathcal{O}_n\,|\,\theta^\mu)$ under the policy $\theta^\mu$, with a hyperbolic tangent activation in the output layer~\cite{haarnoja2018soft}.
We then introduce Gaussian exploration noise during training by sampling $z_n \sim \mathcal{N}\!\left(\mu(\mathcal{O}_n\,|\,\theta^\mu),\sigma^2\right)$, where $\mathcal{N}$ denotes a normal distribution with mean $\mu$ and variance $\sigma^2$, to encourage the agent to explore different actions.
Here $\sigma$ controls the exploration strength.

The noisy output of the actor network is clipped and mapped onto the prescribed action range $[\delta t_{\min},\delta t_{\max}]$ as
\begin{align}
\mathcal{A}_n = \delta t_n
&=
(\delta t_{\max}-\delta t_{\min})
\frac{\operatorname{clip}(z_n,-1,1)+1}{2}
+\delta t_{\min}.
\label{eq:exploration_noise}
\end{align}
Here, $\operatorname{clip}(z_n,-1,1)=\min\{1,\max\{-1,z_n\}\}$ restricts the noisy output to $[-1,1]$, and a subsequent affine transformation rescales it to the physically allowed range $[\delta t_{\min},\delta t_{\max}]$.
This construction ensures that the selected Trotter step size always remains within the prescribed range.
The clipping step maps out-of-range actions to the boundary values, which can lead to an accumulation of probability mass at the edges of the action space~\cite{fujita2018clipped}.

After the agent takes the action $\delta t_n$, the quantum state evolves according to
\begin{align}
|\psi_{n+1}(t_{n+1})\rangle = U(\delta t_n)|\psi_n(t_n)\rangle.
\end{align}
After the propagation, the new observables are measured, from which $\mathcal{O}_{n+1}$ is obtained.
The environment then assigns an immediate reward 
\begin{align}
\mathcal{R}_n =
\Bigl[
(1-\alpha)e^{-\beta_1(\mathcal{E}_{n+1}-\mathcal{E}_0)^2}
+\alpha e^{-\beta_2(\delta\mathcal{E}_{n+1}^2-\delta\mathcal{E}_0^2)^2}
\Bigr] \times
\tanh(\eta\delta t_n),
\end{align}
to evaluate the quality of the Trotter step at the current decision point. 
The updated observation $\mathcal{O}_{n+1}$ is then used as the input for the next step.

In all, we are solving a sequential decision-making problem, with the goal of maximizing the cumulative reward over the entire Trotterization process. The sequence can be described as follows:
\begin{align}
\mathcal{O}_0, d_0
\xrightarrow{\textrm{actor network}}
\mathcal{A}_0 
\xrightarrow{\textrm{time evolution}} 
\mathcal{O}_1, \mathcal{R}_0,d_1   
\cdots 
\xrightarrow{\textrm{actor network}} 
\mathcal{A}_{N-1}
\xrightarrow{\textrm{time evolution}} 
\mathcal{O}_N, \mathcal{R}_{N-1}.\
\end{align}
An episode consists of $N$ sequential Trotter decisions. $d_n$ represents the termination signal, which is defined as
\begin{align}
d_n=
\begin{cases}
1,& n=N-1,\\
0,& \text{otherwise}.
\end{cases}
\end{align}
When the termination signal $d_n=1$, the episode ends and the environment is reset to the initial observation $\mathcal{O}_0$ for the next episode.

The goal of the RL agent is not to maximize the immediate reward $\mathcal{R}_n$ at each step, but to maximize the expected discounted cumulative reward over the whole episode. 
To quantify the long-term value of a decision (i.e., RL action $\mathcal{A}_n$), we define the return from step $n$ as the discounted cumulative reward collected from step $n$ onward,
\begin{align}
G_n
=
\sum_{k=n}^{N-1}
\gamma^{k-n}\mathcal{R}_k,
\end{align}
where $\gamma\in[0,1)$ is the discount factor. The immediate reward $\mathcal{R}_n$ evaluates the quality of the selected time step at the current decision point, whereas the return $G_n$ measures the cumulative effect of this decision through the subsequent Trotterization trajectory. 

The learning objective is to find a policy that maximizes the expected return from the initial observation,
\begin{align}
J=\mathbb{E}[G_0].
\end{align}
Here, the expectation is taken over the episodes generated by the policy during training.
In practice, this expected return is generally unknown and depends on the future rewards.
Thus, DDPG uses a critic network to estimate it through the action value function,
\begin{align}
Q(\mathcal{O}_n,\mathcal{A}_n)
=
\mathbb{E}
\left[
G_n
\mid
\mathcal{O}_n,\mathcal{A}_n
\right].
\end{align}
{It quantifies the expected discounted return starting from the observation $\mathcal{O}_n$, when the agent first takes the action $\mathcal{A}_n$ and then follows the policy $\theta^\mu$ for all subsequent Trotter steps until the end of one episode.}
The actor network is then optimized to choose the action $\mathcal{A}_n=\delta t_n$ that maximizes the critic-estimated action value. 

The data used to update the agent is collected as follows.
At each step, the MDP generates a transition tuple
\begin{align}
\mathcal{D} = \{ (\mathcal{O}_n,\mathcal{A}_n,\mathcal{O}_{n+1},\mathcal{R}_n, d_n ) \}.
\label{seq:transition_data}
\end{align}
These samples are stored in a replay buffer and used to update the actor and critic networks.

When we evaluate the policy, we turn off exploration and use the deterministic actions, i.e., 
\begin{align}
\delta t_n
=
(\delta t_{\max}-\delta t_{\min})
\frac{\mu(\mathcal{O}_n\,|\,\theta^\mu)+1}{2}
+\delta t_{\min}.
\end{align}

The detailed algorithm is provided in Algorithm~\ref{algorithm}.

\subsection{DDPG Training Procedure for RL-Trotter}

{As shown in the previous section,}
DDPG employs two networks: an actor $\mu(\mathcal{O}\,|\,\theta^\mu)$ and a critic $Q(\mathcal{O},\mathcal{A}\,|\,\theta^Q)$.
The actor maps a state directly to a deterministic action, and the critic estimates the action value function.
Here, $\mathcal{A}_n=\delta t_n$ denotes the physical Trotter step size; for notational simplicity, $\mu$ and $\mu'$ throughout this subsection include the affine transformation from the normalized network outputs to physical Trotter step sizes.
{To stabilize training, DDPG introduces target actor and critic networks, $\mu'(\mathcal{O}\,|\,\theta^{\mu'})$ and $Q'(\mathcal{O},\mathcal{A}\,|\,\theta^{Q'})$, whose parameters are slowly updated from those of the online networks, $\mu(\mathcal{O}\,|\,\theta^\mu)$ and $Q(\mathcal{O},\mathcal{A}\,|\,\theta^Q)$. The online networks are directly optimized during training.}
In this section, we describe the training procedure for the four networks in detail~\cite{lillicrap2015continuous}.

During training, mini-batches of the transitions $\mathcal{D}$ {with size $B$} are sampled uniformly from the replay buffer. 
This decorrelates the data and allows DDPG to reuse the past experience efficiently.

Given a mini-batch $\{(\mathcal{O}_i,\mathcal{A}_i,\mathcal{R}_i,\mathcal{O}_{i+1},d_i)\}_{i=1}^{B}$, the target for the critic is constructed from the Bellman equation based on the target critic network,
\begin{align}
y_i
=
\mathcal{R}_i
+
\gamma(1-d_i)
Q'\!\left(
\mathcal{O}_{i+1},
\mu'(\mathcal{O}_{i+1}\,|\,\theta^{\mu'})
\,\middle|\,
\theta^{Q'}
\right).
\label{eq:critic_target}
\end{align}
The critic is then trained by minimizing the mean-squared Bellman error,
\begin{align}
L(\theta^Q)
=
\frac{1}{B}
\sum_{i=1}^{B}
\left[
y_i-Q(\mathcal{O}_i,\mathcal{A}_i\,|\,\theta^Q)
\right]^2.
\label{eq:critic_loss}
\end{align}
Once the critic has been updated,
the actor is updated by maximizing the critic-evaluated return of the current deterministic policy. 
For a mini-batch $\{\mathcal{O}_i\}_{i=1}^B$, the actor objective is
\begin{align}
J(\theta^\mu)
=
\frac{1}{B}\sum_{i=1}^B
Q\!\left(\mathcal{O}_i,
\mu(\mathcal{O}_i\,|\,\theta^\mu)
\,|\,\theta^Q\right).
\end{align}
In implementation, this is equivalently written as minimizing
\begin{align}
L_{\mathrm{actor}}(\theta^\mu)
=
-
\frac{1}{B}\sum_{i=1}^B
Q\!\left(\mathcal{O}_i,\mu(\mathcal{O}_i\,|\,\theta^\mu)\,|\,\theta^Q\right).
\end{align}
The corresponding deterministic policy gradient is
\begin{align}
\nabla_{\theta^\mu} J
\approx
\frac{1}{B}\sum_{i=1}^B
\nabla_{\mathcal{A}}Q(\mathcal{O},\mathcal{A}\,|\,\theta^Q)
\bigg|_{\substack{
\mathcal{O}=\mathcal{O}_i\\
\mathcal{A}=\mu(\mathcal{O}_i\,|\,\theta^\mu)
}}
\nabla_{\theta^\mu}\mu(\mathcal{O}_i\,|\,\theta^\mu).
\label{eq:policy_gradient}
\end{align}
In practice, the actor and critic parameters are both optimized by backpropagation using Adam~\cite{kingma2014adam}, with gradient norm clipping for stability~\cite{pascanu2013difficulty}.
The target networks are then softly updated according to
\begin{align}
\theta^{Q'} &\gets \tau\theta^Q + (1-\tau)\theta^{Q'},\\
\theta^{\mu'} &\gets \tau\theta^\mu + (1-\tau)\theta^{\mu'},
\label{eq:soft_update}
\end{align}
where $\tau\ll 1$ is the soft-update rate. 
These slowly varying target networks reduce oscillations and improve convergence.
The hyperparameters used in the training are shown in Table~\ref{tab:DDPG_parameters}.
The full training procedure is summarized in Algorithm~\ref{algorithm}.

\begin{figure}[t!]
	\centering
	\includegraphics[width=0.8\textwidth]{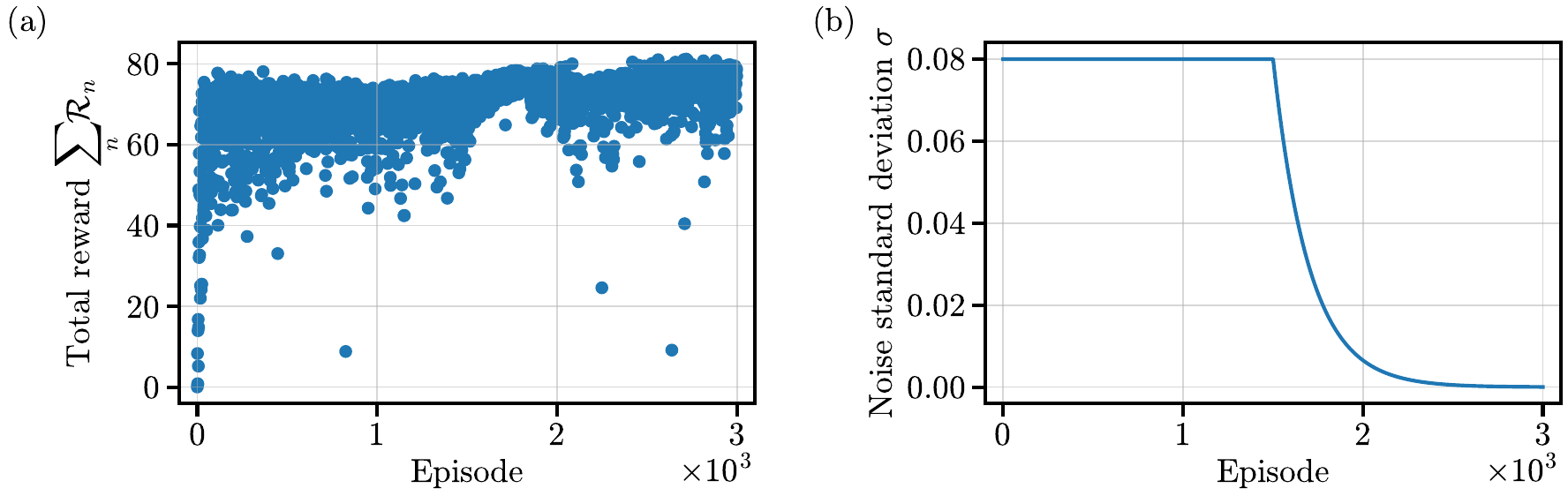}
	\caption{
\textbf{Training results of the DDPG agent for generalization to different parameter settings.}
(a) Total reward as a function of the training episode, showing the learning progress of the agent.
(b) Noise standard deviation $\sigma$ as a function of the training episode, which determines the exploration strength.
The final trained agent is then directly used to evaluate its generalization ability to other quench parameters.
	}
\label{training_results}
\end{figure}

\newpage 
\begin{algorithm}[H]
    \centering
    \caption{DDPG for RL-Trotter}
    \label{algorithm}
    \begin{algorithmic}[1]
        \State Initialize actor $\mu(\mathcal{O}\,|\,\theta^{\mu})$ and critic $Q(\mathcal{O},\mathcal{A}\,|\,\theta^{Q})$
        \State Initialize target networks $\mu'$ and $Q'$ with $\theta^{\mu'}\gets\theta^{\mu}$ and $\theta^{Q'}\gets\theta^{Q}$
        \State Initialize replay buffer $\mathcal{D}$ with capacity $\mathcal{M}$
        \State Set mini-batch size $B$, discount factor $\gamma$, soft-update rate $\tau$, and noise strength $\sigma$

        \For{episode $=1$ to $N_{\mathrm{ep}}$}
            \State Reset environment and prepare the initial state $\ket{\psi_0}$
            \State Measure observables and set $\mathcal{O}_0=\{\mathcal{E}_{n=0},\mathcal{E}_0,\delta\mathcal{E}_{n=0}^2,\delta\mathcal{E}_0^2,t_0=0\}$

            \For{$n=0$ to $N-1$}
            \State Sample exploration noise $z_n \sim \mathcal{N}\!\left(\mu(\mathcal{O}_n\,|\,\theta^\mu),\sigma^2\right)$
            \State Select action $\delta t_n =
            (\delta t_{\max}-\delta t_{\min})
            \frac{\operatorname{clip}(z_n,-1,1)+1}{2}
            +\delta t_{\min}.$
                \State Evolve the wavefunction $\ket{\psi_{n+1}}\gets U(\delta t_n)\ket{\psi_n}$
                \State Update accumulated time $t_{n+1}\gets t_n+\delta t_n$
                \State Measure observables and construct next state $\mathcal{O}_{n+1}$
                \State Compute reward $\mathcal{R}_n$ from the measured observables
                \State Set terminal flag $d_n\gets 1$ if $n=N-1$, otherwise $d_n\gets 0$
                \State Store transition $(\mathcal{O}_n,\delta t_n,\mathcal{R}_n,\mathcal{O}_{n+1},d_n)$ in $\mathcal{D}$

                \If{$|\mathcal{D}|\ge B$}
                    \State Sample mini-batch $\{(\mathcal{O}_i,\mathcal{A}_i,\mathcal{R}_i,\mathcal{O}_{i+1},d_i)\}_{i=1}^{B}$ from $\mathcal{D}$
                    \State Compute target values
                    \begin{align*}
                        y_i \gets \mathcal{R}_i+\gamma(1-d_i)\,
                        Q'\!\left(
                        \mathcal{O}_{i+1},
                        \mu'(\mathcal{O}_{i+1}\,|\,\theta^{\mu'})
                        \,\middle|\,
                        \theta^{Q'}
                        \right)
                    \end{align*}
                    \State Update critic by minimizing
                    \begin{align*}
                        L=\frac{1}{B}\sum_{i=1}^{B}
                        \left[
                        y_i-Q(\mathcal{O}_i,\mathcal{A}_i\,|\,\theta^Q)
                        \right]^2
                    \end{align*}
                    \State Update actor using the deterministic policy gradient in Eq.~\eqref{eq:policy_gradient}
                    \State Soft-update target critic:
                    \begin{align*}
                        \theta^{Q'}\gets\tau\theta^Q+(1-\tau)\theta^{Q'}
                    \end{align*}
                    \State Soft-update target actor:
                    \begin{align*}
                        \theta^{\mu'}\gets\tau\theta^{\mu}+(1-\tau)\theta^{\mu'}
                    \end{align*}
                \EndIf

                \State Set $\mathcal{O}_n\gets\mathcal{O}_{n+1}$ and $\ket{\psi_n}\gets\ket{\psi_{n+1}}$
            \EndFor
        \EndFor
        \State \textbf{Return} trained policy $\mu(\mathcal{O}\,|\,\theta^{\mu})$
    \end{algorithmic}
\end{algorithm}

\begin{table}[htbp]
\centering
\caption{Hyperparameters used to train the DDPG agent.}
\label{tab:DDPG_parameters}
\small
\renewcommand{\arraystretch}{1.15}
\setlength{\tabcolsep}{8pt}
\begin{tabular}{|c|c|}
\hline
\hline
Actor and critic hidden layers &300,200,100 \\  \hline 
Hidden-layer activation function & ReLU \\ \hline
Number of training episodes & 3000 \\ \hline
Maximum Trotter step $\delta t_{\max}$  & $0.8 J^{-1}$  \\ \hline
Minimum Trotter step $\delta t_{\min}$  & $0.01 J^{-1}$ \\ \hline
Reward parameters &  $\alpha = 0.5$, $\beta_1=20J^{-2}$, $\beta_2 = 20J^{-4}$, $\eta = 5J$\\ \hline
Discount factor $\gamma$ &  $0.95$\\ \hline
Soft-update rate $\tau$ &       $0.001$  \\   \hline
Replay-buffer capacity $M$ &          $50000$ \\ \hline
Mini-batch size $B$ & $256$ \\ \hline
Actor learning rate & $0.0005$ \\   \hline
Critic learning rate & $0.001$ \\  \hline
gradient norm clipping & $1.0$ \\ \hline
\hline
\end{tabular}
\end{table}

\subsection{Training details for the agent used to demonstrate generalization across different parameter settings}
\label{sec:training_details}

Here, we present the training details of the agent used in the main text that demonstrates the generalization capability across different initial states.
Both the actor and critic are implemented as fully connected neural networks with three hidden layers containing 300, 200, and 100 neurons, respectively. The remaining training parameters are summarized in Table~\ref{tab:DDPG_parameters}.

The training results are shown in Fig.~\ref{training_results}. 
{Benefiting from the compact neural-network architecture, the agent can be trained efficiently within only $3000$ episodes.}
Figure~\ref{training_results}(a) shows the total reward per episode, which characterizes the learning progress of the agent. 
The reward increases rapidly during the early stage of training and gradually converges.
During the final stage, the episode reward becomes stable, with an average value of $74.50$ over the last $100$ episodes.

Figure~\ref{training_results}(b) shows the standard deviation $\sigma$ of the exploration noise as a function of the training episode. The noise strength is gradually reduced during training, allowing the agent to explore the action space more broadly at the early stage and to increasingly exploit the learned policy at the later stage. The final trained agent is then used to evaluate its generalization to other quench parameters.

\begin{figure}[tbh]
	\centering
	\includegraphics[width=0.8\textwidth]{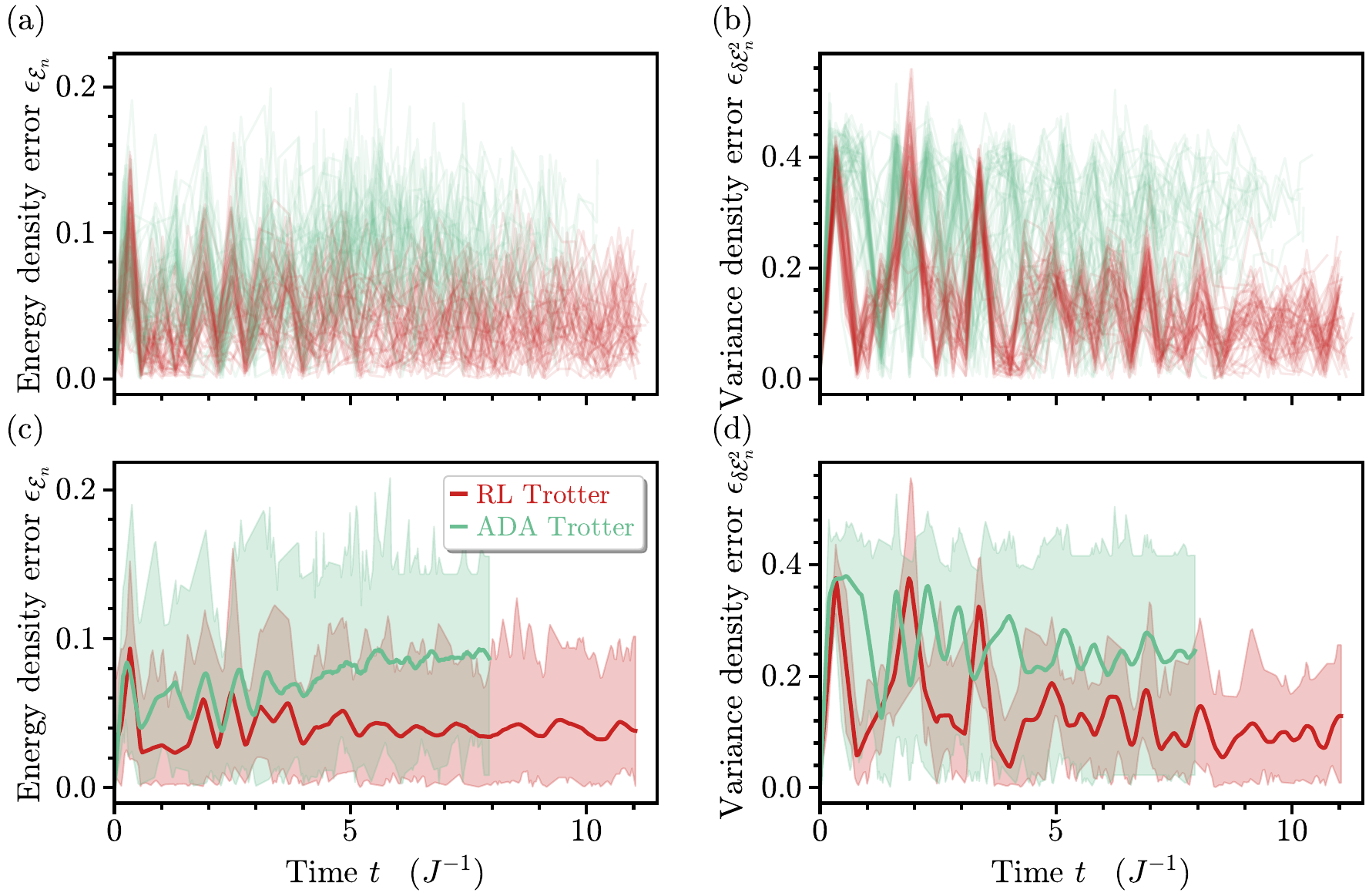}
	\caption{
\textbf{Robustness of RL-Trotter and ADA-Trotter under measurement noise.}
(a,b) Time evolution of the energy density error $\epsilon_ \mathcal{E}$ and the energy variance-density error $\epsilon_{ \delta \mathcal{E}^2}$ for different noise realizations. Red and green curves correspond to RL-Trotter and ADA-Trotter, respectively. We use $50$ different noise realizations, and the noise is sampled from a normal distribution with standard deviation $\sigma_1 = \sigma_2 =0.03$. The total evolution time for RL-Trotter is $t = 11.04 \pm 0.09 J^{-1}$ and for ADA-Trotter is $t = 7.94 \pm 1.42 J^{-1}$.
(c,d) The same data after linearly interpolating all trajectories onto a common time grid. The range of the time grid is chosen as $[0,7.94]$ for ADA-Trotter and $[0,11.04]$ for RL-Trotter.
The solid curves denote the mean trajectories over different noise realizations, while the shaded regions indicate the corresponding pointwise min-max envelopes.
Compared with ADA-Trotter, RL-Trotter reaches longer evolution times and shows smaller errors in both conserved quantities.
    }
\label{SI_noise}
\end{figure}

\subsection{Description of supplementary videos---detailed dynamics during training}
\label{sec:supp_videos}

This work is accompanied by $6$ supplementary videos illustrating representative stages of the training process.
We consider the deterministic quench process discussed in Sec.~\ref{Sec:Benchmarking_RL} of the main text.
The corresponding evolution of the total reward during training is shown in Fig.~\ref{fig1_illustration}(b).
Here, we present the detailed dynamics of the energy density and the energy variance density in several representative training episodes.
For each video, the sequence of Trotter step sizes is generated online during the corresponding training episode and therefore includes the exploration noise added to the actor output. Note that the agent continues to update within each episode.

\textit{episode\_0.mp4}.---
At the beginning of training, both the actor and critic networks are initialized using the Kaiming uniform initialization~\cite{he2015delving}. Specifically, the weights of the hidden fully connected layers are initialized with a negative-slope parameter $a{=}0.1$, while all hidden-layer biases are initialized to zero. 
The weights and biases of the output layers are sampled uniformly from $[-3\times10^{-3},3\times10^{-3}]$.
The selected Trotter step size rapidly approaches $\delta t{=}0.8$, corresponding to the upper bound of the predefined action range. 
Consequently, the evolution proceeds rapidly in physical time, but both the energy density and energy variance density deviate strongly from their target values. 
The results indicate that the initially randomized agent has not yet learned an effective Trotter policy.

\textit{episode\_300.mp4}.---
After approximately $300$ training episodes, the Trotter step size fluctuates mainly within an intermediate range, approximately $0.2 \lesssim \delta t \lesssim 0.3$. 
This already represents a substantial qualitative change from the initial behavior.
More importantly, a primitive self-correcting behavior starts to emerge. 
During the early part of the evolution, the deviation of the energy density from its target value gradually increases. 
After approximately $20$ Trotter steps, however, the trend reverses and the energy density begins to move back toward the target value, only to drift away again after approximately $50$ steps.
This behavior indicates that the agent is beginning to exploit subsequent actions to compensate for errors accumulated at earlier steps, although the correction remains incomplete and the energy variance density still exhibits a noticeable long-time drift.

\textit{episode\_560.mp4}.---
In this episode, the learned policy is qualitatively different from the \textit{episode\_300.mp4} case.
During the first part of the evolution, the agent again drives the Trotter step size toward the upper bound, $\delta t\simeq0.8$. 
This produces large deviations in both the energy density and the energy variance density.
After roughly $30$ Trotter steps, the policy undergoes an abrupt change: the selected step size rapidly collapses from values close to the upper bound to values close to the lower bound $\delta t=0.01$. 
The subsequent evolution therefore proceeds only very slowly in physical time.
This behavior can be interpreted as an immature error-control strategy in which the agent reacts to the previously accumulated error by strongly suppressing all subsequent step sizes.
Although this limits further error accumulation, it also prevents the simulation from efficiently extending the accessible evolution time. 

\textit{episode\_610.mp4}.---
About $50$ training episodes later, the agent continuously adjusts the Trotter step size within an intermediate range, approximately $0.1 \lesssim \delta t \lesssim 0.35$.
Correspondingly, the energy density and energy variance density remain substantially closer to their target values than in \textit{episode\_560.mp4}. 
The step-size sequence exhibits frequent local adjustments rather than a single abrupt switch, suggesting that the agent has started to use the instantaneous error information to perform continuous feedback control throughout the evolution.
Nevertheless, noticeable long-time deviations remain, indicating that the policy has not yet fully converged.

\textit{episode\_1600.mp4}.---
In this episode, both the energy density and the energy variance density remain close to their target values over a much longer part of the evolution. 
In particular, the large deviations observed in the earlier episodes are strongly suppressed.
The selected Trotter step size also develops a systematic temporal structure. 
The agent initially chooses relatively large steps, with $\delta t$ reaching approximately $[0.3,0.35]$, and gradually reduces the step size as the evolution proceeds. 
During the late-time dynamics, $\delta t$ decreases toward values of order $0.1$. Such behavior demonstrates that the agent has learned to use comparatively large steps when the accumulated error remains small, while progressively reducing the step size when tighter control is required at later times.

\textit{episode\_2852.mp4}.---
After approximately $2850$ training episodes, the agent exhibits a well-developed adaptive policy. 
The energy density and energy variance density remain confined close to their corresponding target values throughout the evolution. The improved accuracy is not achieved simply by continuously reducing the Trotter step size. 
After an initial transient, the agent is able to maintain a relatively large step size, approximately $0.32 \lesssim \delta t \lesssim 0.36$, over an extended part of the evolution. 
The step size is subsequently adjusted smoothly rather than collapsing toward zero. 
As a result, the total evolution time is larger than that in \textit{episode\_1600.mp4} while the deviations of the conserved quantities remain controlled.

\subsection{Computational resource discussion}
The training simulations were performed on a single NVIDIA A100 PCIe GPU with $40$ GiB of device memory or a single NVIDIA H200 NVL GPU with approximately $140$ GiB of device memory. The calculations were implemented using PyTorch 2.8.0~\cite{paszke2019pytorch} with CUDA 12.6 support. In both cases, a full GPU device without MIG partitioning was used. 

The classical emulation during training was allocated $8$ CPU cores and $64$ GB of system memory, using the package QuSpin 1.0.0~\cite{weinberg2017quspin,weinberg2019quspin}. 
The entire training process, including the simulation of quantum dynamics, is efficient. It can be completed within approximately $2$ hours.

\subsection{Motivation for the reward design}
\label{sec:RL_reward}

In our work, the reward is designed to favor large time steps while penalizing errors in the conserved quantities. The explicit form reads
\begin{align}
\mathcal{R}_n =
\Bigl[
(1-\alpha)e^{-\beta_1(\mathcal{E}_{n+1}-\mathcal{E}_0)^2}
+\alpha e^{-\beta_2(\delta\mathcal{E}_{n+1}^2-\delta\mathcal{E}_0^2)^2}
\Bigr] \times
\tanh(\eta\delta t_n). \label{seq:RL_reward}
\end{align}
The factor $\tanh(\eta\delta t_n)$ rewards larger steps, whereas the exponential terms suppress actions that generate excessive deviations in the energy density or energy-variance density. 
The optimal policy therefore seeks the longest simulation time while keeping overall errors in conserved quantities self-correcting. 
Here, we discuss several considerations underlying the reward design, which may also provide useful guidance for constructing reward functions for other optimization targets.

State fidelity of the evolved state with respect to the exact state at the same time step, or the fidelity of the final evolved state with respect to the target state, is a natural choice for the reward function in quantum control problems~\cite{Bolens2021Reinforcement}.
However, this choice is not suitable in RL-Trotter, for several reasons.

First, small Trotter step sizes usually lead to high fidelity. As a result, a fidelity-based reward may bias the agent toward choosing extremely small $\delta t_n$ at every Trotter step. 
Although this suppresses the local Trotter error, it also advances the physical evolution very slowly and therefore makes the simulation inefficient.

Second, the fidelity is not a suitable measure at long evolution times or for large system sizes, since it can decay to very small values even when local observables are accurately reproduced. This makes it difficult to assess the quality of the evolved state based solely on the global fidelity.
Possible alternatives include local measures, such as the fidelity between reduced density matrices. However, the main problem is still the fact that exact local information is inaccessible for large systems.

Third, assigning the reward only according to the fidelity at the end of the entire evolution may be ineffective for an off-policy algorithm such as DDPG. 
In DDPG, as discussed later, transitions are stored as tuples in a replay buffer, from which minibatches are sampled during training. 
If only a final reward is provided, most transitions will carry zero reward, making it difficult for the agent to learn an effective policy.

The reward function in Eq.~\eqref{seq:RL_reward} has several advantages.

First, the multiplicative structure couples the error-dependent term to the step-size-dependent term, such that a large reward can be obtained only when the agent simultaneously maintains small Trotter errors and selects a sufficiently large step size. This balance directly reflects the physical objective of the optimization.
For small $\delta t_n$, the function $\tanh(\eta\delta t_n)$ increases approximately linearly, providing a strong incentive for the agent to increase the step size. 
As $\delta t_n$ becomes larger, however, the $\tanh$ function gradually saturates, and the marginal reward gained by further increasing the step size becomes small. 
The reward is then increasingly governed by the error-dependent term, which penalizes the growth of Trotter errors and prevents the agent from choosing excessively large step sizes.

Second, the observables entering the reward function are experimentally accessible. In our case, the energy and its variance can be estimated directly from measurements of the quantum state, without requiring full-state tomography. This makes the reward compatible with realistic quantum experiments and allows the agent to receive feedback using only a limited set of physically measurable quantities.

Third, the critic network is much easier to train when nearby states and actions yield similar returns, i.e., the reward is continuous. This smoothness in the reward landscape improves stability and sample efficiency.

\section{Supplemental results for the comparison between different Trotter algorithms}

{Here we provide additional results to demonstrate the key advantages of RL-Trotter.}

\subsection{Robustness comparison between RL-Trotter and ADA-Trotter under measurement noise}
\label{sec:Robustness_comparison}

We compare the robustness of RL-Trotter and ADA-Trotter under measurement noise. 
For each Trotterization algorithm, we perform $50$ independent simulations. Each simulation consists of $40$ Trotter steps, with measurement-noise strengths $\sigma_1=\sigma_2=0.03$.
We show the average error and its standard deviation in the main text and the detailed trajectories in Fig.~\ref{SI_noise}.

We first compare the total evolution time. We find that ADA-Trotter is more sensitive to measurement noise than RL-Trotter.
The average total time of ADA-Trotter is $t=7.94J^{-1}$, which is significantly shorter than that achieved by RL-Trotter, $t=11.04J^{-1}$.
The larger standard deviation of evolution time for ADA-Trotter (i.e., $1.42J^{-1}$) shows that its performance changes more strongly from one noise realization to another.
In contrast, RL-Trotter shows a much smaller fluctuation ($0.09J^{-1}$), indicating stronger robustness against measurement noise.

RL-Trotter not only extends the average accessible evolution time but also accurately reproduces the expectation values of physical observables.
We further compare the time evolution of the energy density error $\epsilon_\mathcal{E}$ and the energy variance density error $\epsilon_{\delta \mathcal{E}^2}$.
To average over different noise realizations, all trajectories are linearly interpolated onto a common time grid for each method.
The common time grid is chosen over the corresponding average evolution-time interval, namely $[0,11.04]$ for RL-Trotter and $[0,7.94]$ for ADA-Trotter.
The solid curves denote the mean trajectories over different noise realizations, while the shaded regions indicate the corresponding pointwise min-max envelopes.
RL-Trotter exhibits smaller average errors and narrower fluctuation ranges in both conserved quantities compared with ADA-Trotter, further demonstrating its robustness against measurement noise.

\begin{figure}[t!]
	\centering
	\includegraphics[width=0.8\textwidth]{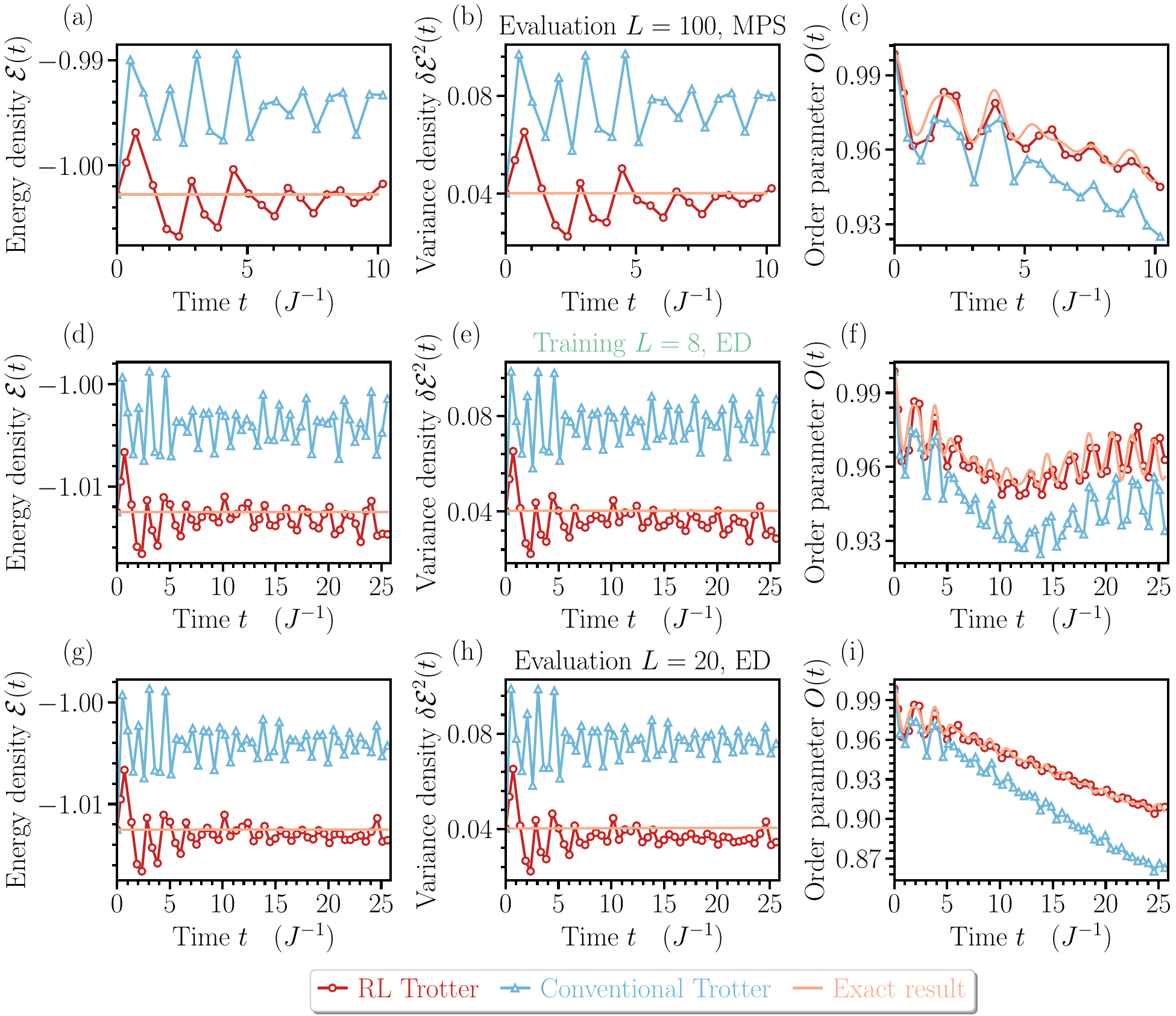}
\caption{ \textbf{MPS benchmark of the RL-Trotter protocol for $L=100$.} Time evolution of (a) the energy density $\mathcal{E}(t)$, (b) the energy-variance density $\delta \mathcal{E}^2(t)$, and (c) the staggered order parameter $O(t)$ for a system with $L=100$. 
The RL agent is trained on small systems with $L=8$ using ED and is directly applied to the large system using MPS representation. 
The quench is performed from $h_z=0$ and $h_x=0.1$ to $h_z=0.3$ and $h_x=0.3$. 
The red circles denote the RL-Trotter results, the blue triangles denote the conventional Trotter results with the same total evolution time and number of Trotter steps, and the orange solid curve denotes the exact result. 
The RL-Trotter protocol preserves both the energy and energy variance (with errors $\langle \epsilon_\mathcal{E}\rangle = 1.62 \times 10^{-3}J$ and $\langle \epsilon_{\delta \mathcal{E}^2}\rangle = 7.07 \times 10^{-3}J^{2}$) much more accurately than the conventional Trotter method, and also reproduces the dynamics of the staggered order parameter with smaller deviations.
(d-f) The corresponding results for a system with $L=8$ (green title), which is the system size used for training.
(g-i) The corresponding results for a system with $L=20$, which is the system size used for evaluation in the main text.
The results are obtained from the ED method.
The RL agent is able to capture the different dynamical behaviors of the order parameter $O(t)$ for both system sizes, demonstrating its generalization ability.
}
\label{MPS_Observables_compare}
\end{figure}

\subsection{Generalization of conventional Trotter and RL-Trotter to $L=100$ via MPS}
\label{sec:Generalization_to_large_systems_via_MPS}

In this section, we test the effectiveness of the RL agent trained on small systems with $L=8$ and then applied to larger systems with $L=100$ based on matrix product state (MPS) representation~\cite{schollwock2011density}.

To simplify the numerical simulation while still capturing the essential physics, we consider a quench from $h_z=0$ and $h_x=0.1$ to $h_z=0.3$ and $h_x=0.3$. 
For this quench, the maximum bond dimension grows much more slowly than in quenches that cross the phase boundary. 
We set the maximum bond dimension to $\chi_{\max}=50$, which is enough to capture the dynamics.
The ground state of the initial Hamiltonian is obtained using the density matrix renormalization group method
and the real-time evolution is performed using the two-site time-dependent variational principle {(TDVP)}. 
The simulation is carried out with the TeNPy package~\cite{hauschild2018efficient}.

We train the RL agent on systems with $L=8$ using ED and then directly apply the trained agent to the system with $L=100$ simulated by {TDVP}. 
Periodic boundary conditions are used during training, while open boundary conditions are used during evaluation. 
We also compare the RL-Trotter results with those obtained from the conventional Trotter method using the same total evolution time and the same number of Trotter steps. 
The results are shown in Fig.~\ref{MPS_Observables_compare}.

We find that the errors in both the energy density and the energy-variance density are still corrected by the RL agent. 
The corresponding average errors are small, $\langle \epsilon_\mathcal{E}\rangle = 1.62 \times 10^{-3}J$ and $\langle \epsilon_{\delta \mathcal{E}^2}\rangle = 7.07 \times 10^{-3}J^{2}$, respectively. 
In contrast, the conventional Trotter method shows much larger deviations in both quantities, with average errors of $\langle \epsilon_\mathcal{E}\rangle =  8.15 \times 10^{-3}J$ and $\langle \epsilon_{\delta \mathcal{E}^2}\rangle = 3.46 \times 10^{-2}J^{2}$, respectively. 

The conservation of the energy and energy variance also leads to better accuracy for other observables. 
We consider the time evolution of the staggered magnetization $O(t)$ along the $z$ direction, as shown in Fig.~\ref{MPS_Observables_compare}(c). 
The RL agent accurately reproduces the time evolution of the staggered magnetization, while the conventional Trotter method shows clear deviations. 
The corresponding average error is $\langle \epsilon_O \rangle =  3.47 \times 10^{-3}$ for the RL agent, compared with $\langle \epsilon_O \rangle =  1.39 \times 10^{-2}$ for the conventional Trotter method.

{In all, we find that all error measures have been suppressed by around 1 order of magnitude.}

As a comparison, we also test the RL agent on a system with $L = 8$, which is the system size that we used for training, and on a system with $L=20$, which we have also used for evaluation in the main text.
The results are shown in the last two rows of Fig.~\ref{MPS_Observables_compare} respectively.
We find that the order parameter $O(t)$ has different dynamical behaviors for $L = 8$ and $L = 20$. However, the RL agent is able to accurately reproduce the dynamics of $O(t)$ for both system sizes.

\begin{figure}[t!]
	\centering
	\includegraphics[width=0.8\textwidth]{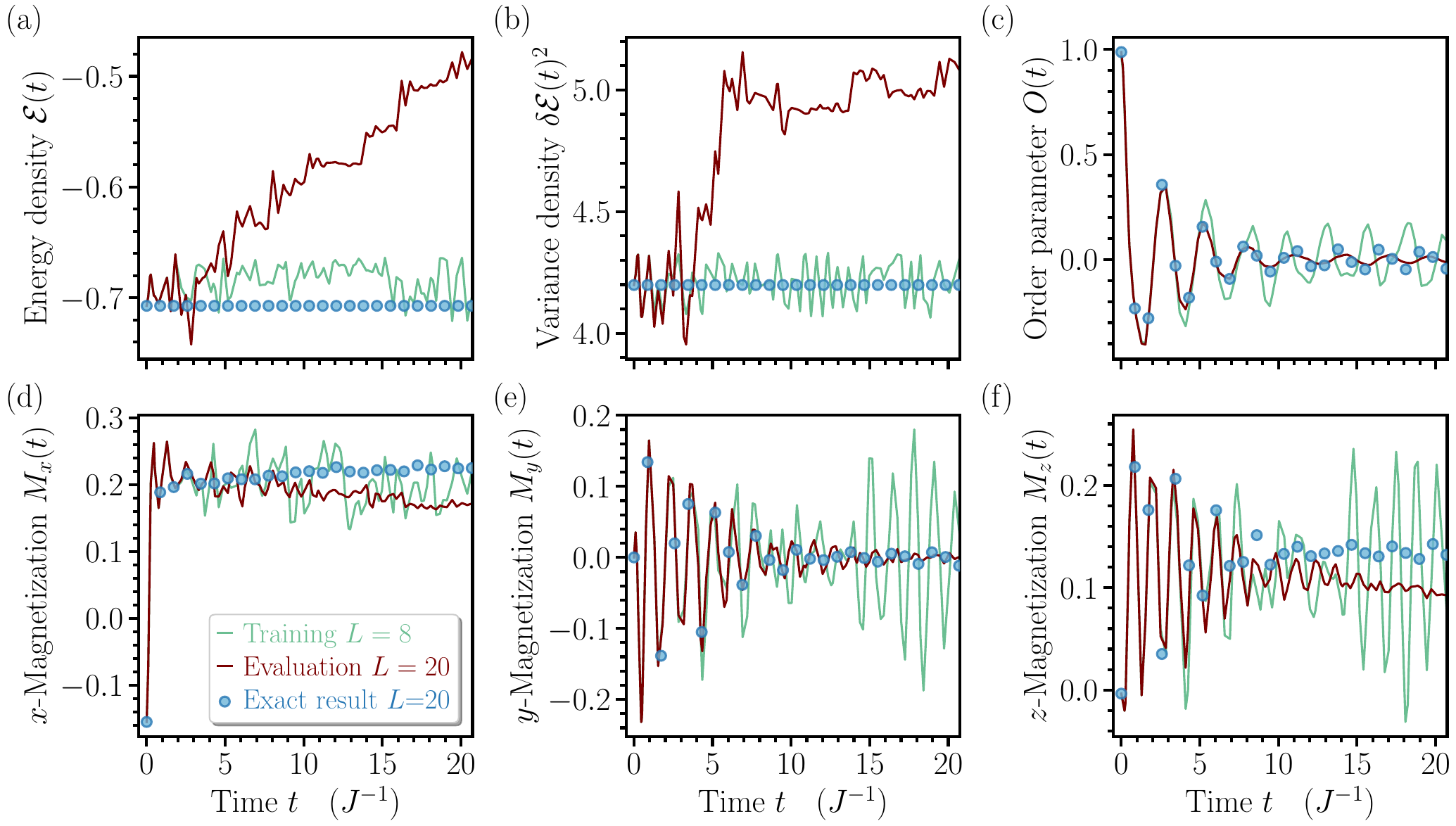}
	\caption{
\textbf{Generalization of ADA-Trotter to larger system sizes.}
The Trotter step-size sequence $\{\delta t_n\}$ is obtained using ADA-Trotter for a system with $L=8$ (green curves) and directly applied to a larger system with $L=20$ (red curves). The exact evolution for $L=20$ is shown by the blue dots. Panels (a) and (b) show that the energy density and energy-variance density deviate from their exact values, indicating that the error constraints optimized for $L=8$ are no longer satisfied at $L=20$. This violation further induces errors in other observables, as shown in the remaining panels. We consider a quench from $h_x=0.3$, $h_z=0.3$ to $h_x=-1.7$, $h_z=-1.3$. For the $L=8$ system, the ADA tolerances are set to $d_{\mathcal{E}}=4.35\times10^{-2}$ and $d_{\delta\mathcal{E}^{2}}=1.36\times10^{-1}$.	
}
\label{ADA_generalization_size}
\end{figure}

\subsection{
Failure of the
Generalization of ADA-Trotter to larger system sizes}
\label{subsec:failure_generalization_ADA}
To demonstrate the advantages of RL-Trotter over ADA-Trotter, we show that the Trotter step size sequence selected by ADA-Trotter cannot be applied to larger system sizes. 

In ADA-Trotter, the feedback loop determines a sequence of Trotter step sizes ${\delta t_n}$ for a specific system, rather than learning a transferable policy $\theta^\mu$.
Therefore, the resulting time-step sequence is tied to the system size for which it is generated. 
When this sequence is directly applied to a larger system, it is no longer guaranteed to satisfy the imposed constraints and can lead to larger Trotter errors. 

The results are shown in Fig.~\ref{ADA_generalization_size}. 
The Trotter step sizes ${\delta t_n}$ are first obtained from ADA-Trotter for a system of size $L=8$ and are then directly applied to a larger system with $L=20$. 
We find significant deviations in both the energy density and the energy variance density from their exact values, indicating that the constraints are no longer satisfied. 
These results show that the time step sequence generated by ADA-Trotter does not generalize well to larger system sizes, thereby highlighting the advantage of the RL-Trotter.

{Whereas the energy and its variance significantly deviate from the exact values, several local observables, e.g., the magnetization in the $y$ direction $M_y$ and the staggered order parameter $O$, still behave well. This behavior can be understood as follows. As $M_y$ and $O$ decay to $0$ eventually, they are not sensitive to the errors in energy and variance. In contrast, the other observables such as the magnetization in the $x$ and $z$ directions show substantial deviations from the exact evolution.}

\subsection{Rewards of the conventional Trotter}
\label{sec:Reward of the conventional Trotter}
In this section, we consider the reward obtained from the conventional Trotter. 
The conventional Trotter with a fixed time step can be viewed as a special case of the policy space. 
We will show that the reward obtained from the conventional Trotter is much smaller than that obtained from the RL-Trotter.

\begin{figure}[tbh]
	\centering
	\includegraphics[width=0.4\textwidth]{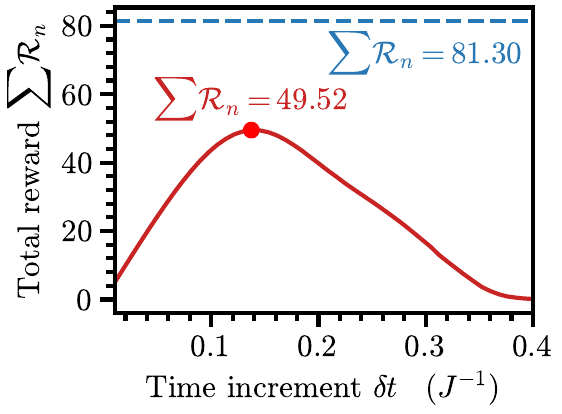}
\caption{
\textbf{Reward comparison between conventional Trotter and RL-Trotter.}
Total reward $\sum_n R_n$ accumulated over the first $100$ Trotter steps for the quench from $h_x=0.3$, $h_z=0.3$ to $h_x=-1.7$, $h_z=-1.3$ in a system with $L=16$. 
The red curve shows the reward obtained by conventional Trotterization with a fixed Trotter step size $\delta t$. 
The red marker denotes the maximum reward among the fixed-step protocols, with $\sum_n R_n=49.52$ at $\delta t=0.14J^{-1}$. 
The blue dashed line indicates the reward obtained by the RL-Trotter protocol, $\sum_n R_n=81.30$, using the same reward function. 
This comparison shows that the learned policy can choose an adaptive sequence of Trotter step sizes that achieves a substantially larger total reward than the best fixed-step protocol considered here. The parameters in the reward function are set to $\beta_1=20J^{-2}$, $\beta_2=20J^{-4}$, $\eta=5J$, and $\alpha=0.5$. 
}
\label{reward_landscape}
\end{figure}

We consider the quench from $h_x=0.3$, $h_z=0.3$ to $h_x=-1.7$, $h_z=-1.3$ for a system with $L=16$, and evaluate the reward over the first $100$ Trotter steps.

The reward obtained by fixed-step Trotterization for different choices of the Trotter step size $\delta t$ is shown in Fig.~\ref{reward_landscape}. Among these fixed-step protocols, the maximum reward is about $49.52$, achieved at $\delta t=0.14 J^{-1}$. This value is much smaller than the reward obtained by the RL-Trotter, which is about $81.30$, as indicated by the blue dashed line. 
Therefore, the learned policy is able to choose a sequence of Trotter step sizes that gives a substantially higher reward than any fixed-step conventional Trotter protocol in this comparison.

It is important to note that the trained agent is not optimized specifically for this single quench process or this particular reward landscape. During training, exploration noise is always included, and the agent learns from a range of trajectories rather than from one fixed quench. 
Thus, although the learned policy remains effective, it may not be optimal for this specific case.
The parameters in the reward function are set to $\beta_1=20J^{-2}$, $\beta_2=20J^{-4}$, $\eta=5J$, and $\alpha=0.5$.

\begin{figure}[t!]
	\centering
	\includegraphics[width=0.7\textwidth]{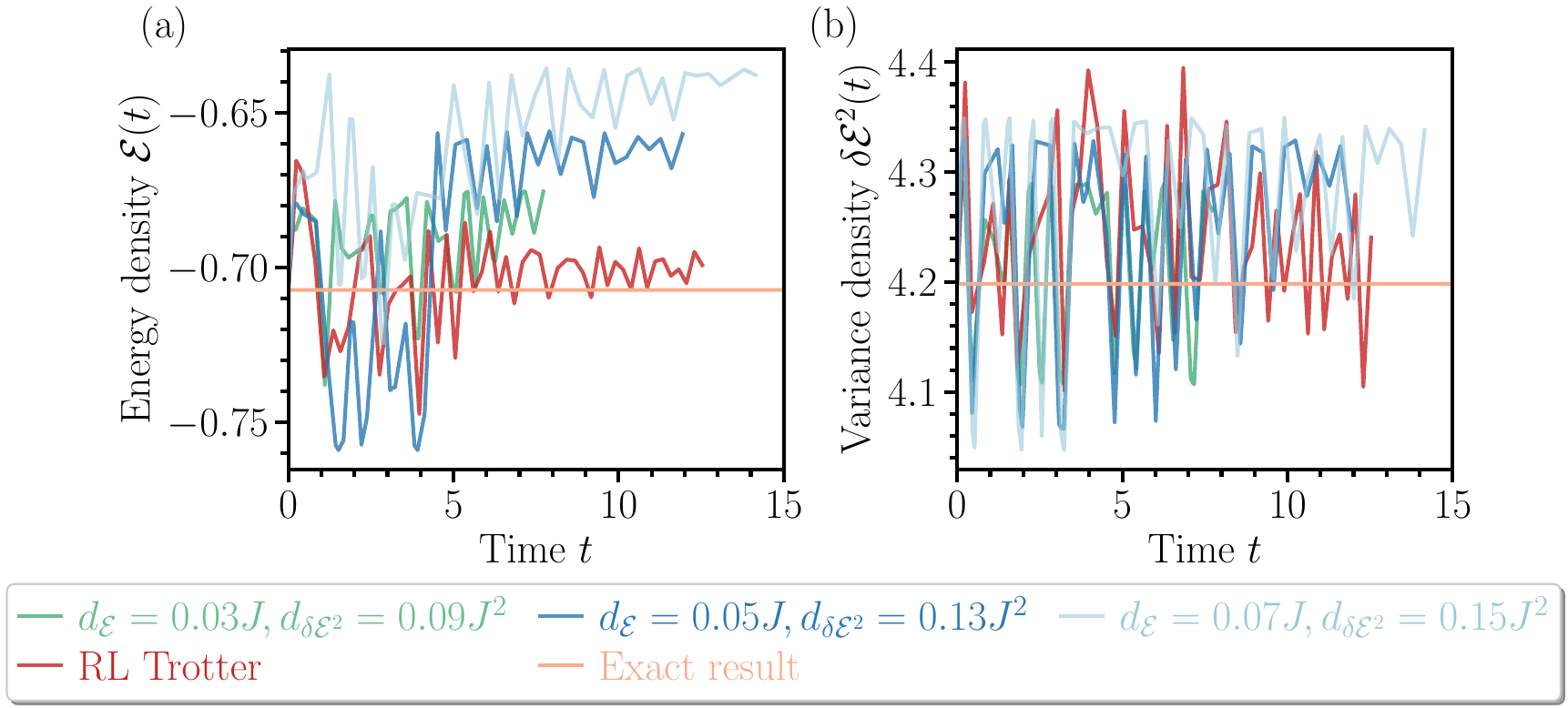}
\caption{
   \textbf{Comparison of ADA-Trotter with different error tolerances and RL-Trotter over the first $50$ Trotter steps.}
(a) Energy density $\mathcal{E}(t)$ and (b) energy-variance density $\delta\mathcal{E}^{2}(t)$ as functions of time $t$.
The green, blue, and cyan curves show ADA-Trotter results for the tolerance sets $(d_{\mathcal{E}},d_{\delta\mathcal{E}^{2}})=(0.03J,0.09J^{2})$, $(0.05J,0.13J^{2})$, and $(0.07J,0.15J^{2})$, respectively.
The red curve shows the RL-Trotter result, while the orange horizontal line denotes the exact value.
The corresponding total evolution times after $50$ steps are $t_{50}=7.72J^{-1}$, $11.94J^{-1}$, and $14.15J^{-1}$ for the three ADA tolerance sets, respectively, compared with $t_{50}=12.53J^{-1}$ for RL-Trotter.
Relaxing the ADA tolerances extends the accessible evolution time but also produces larger deviations in the energy density.
RL-Trotter therefore achieves a more favorable balance between evolution time and simulation accuracy.}
\label{ADA_comparison}
\end{figure}

\subsection{The results of ADA-Trotter with different error tolerances}
\label{subsec:ADA_with_different_tolerance}
In this section, we examine how the predefined constraint tolerances affect the performance of ADA-Trotter.
We consider three different sets of error tolerances and compare the resulting dynamics with those obtained using RL-Trotter. 
The RL agent is identical to that used in Fig.~\ref{quench_dynamics} of the main text. The corresponding energy density and energy-variance density are shown in Figs.~\ref{ADA_comparison}(a) and \ref{ADA_comparison}(b), respectively.

We first consider the tolerance set
$d_{\mathcal{E}}=0.03J$ and
$d_{\delta\mathcal{E}^{2}}=0.09J^{2}$
(green curve), as used in the main text.
Under these restrictive tolerances, ADA-Trotter reaches a total evolution time of only
$t_{50}=7.72J^{-1}$,
substantially shorter than
$t_{50}=12.53J^{-1}$
achieved by RL-Trotter.
Although the energy density and energy-variance density remain close to their exact values, the accessible evolution time is strongly limited by the strict error constraints.

We then relax the tolerances to
$d_{\mathcal{E}}=0.05J$ and
$d_{\delta\mathcal{E}^{2}}=0.13J^{2}$
(blue curve).
The total evolution time increases to
$t_{50}=11.94J^{-1}$,
but remains shorter than that of RL-Trotter.
Meanwhile, the energy density shows noticeably larger deviations from the exact result than in the RL-Trotter evolution.

When the tolerances are further relaxed to
$d_{\mathcal{E}}=0.07J$ and
$d_{\delta\mathcal{E}^{2}}=0.15J^{2}$
(cyan curve), ADA-Trotter reaches
$t_{50}=14.15J^{-1}$,
slightly exceeding the evolution time achieved by RL-Trotter.
This extension, however, comes at the cost of a substantial loss of accuracy.
In particular, the energy density exhibits persistent deviations from the exact value over a large fraction of the evolution.

By contrast, RL-Trotter achieves a long evolution time while keeping both the energy density and the energy-variance density close to their exact values.
This comparison demonstrates that the learned policy achieves a more favorable trade-off between accessible evolution time and simulation accuracy than the local greedy policy employed by ADA-Trotter.

\clearpage

\end{widetext}

\fi 

\ifcombinedmode

\else
\bibliography{bibfile}
\fi 

\end{document}